\documentclass{article}
\usepackage[normalem]{ulem}
\usepackage{geometry}
\usepackage{amsmath,amssymb,bm,amsthm}

\usepackage{bm}

\usepackage{graphicx}
\usepackage{subcaption}
\usepackage{url}

\usepackage{tikz} 
\usetikzlibrary{shapes,arrows,decorations.markings,chains, decorations.pathreplacing,patterns,calc}

\usepackage{algorithm}
\usepackage{algpseudocode}
\usepackage{enumitem}
\makeatletter
\renewcommand{\ALG@beginalgorithmic}{\small}
\makeatother

\usepackage{natbib}
 \bibpunct[, ]{(}{)}{,}{a}{}{,}%
 \def\bibfont{\small}%
\title{\textbf{Managing driving modes in automated driving systems}}
\author{
David Ríos Insua, Institute of Mathematical Sciences (ICMAT-CSIC)\\
William N. Caballero, Air Force Institute of Technology\\
Roi Naveiro, Institute of Mathematical Sciences (ICMAT-CSIC)
\vspace{0.3cm}
}   

\date{~}
\begin{document}
\maketitle

\begin{abstract}
    Current  technologies are unable to produce massively deployable, fully autonomous vehicles that do not require human intervention. Such technological limitations are projected to persist for decades. Therefore, roadway scenarios requiring a driver to regain control of a vehicle, and vice versa, will remain critical to the safe operation of semi-autonomous vehicles for the foreseeable future. Herein, we adopt a comprehensive perspective on this problem taking into account the operational design domain, driver and environment monitoring, trajectory planning, and driver intervention performance assessment. Leveraging decision analysis and Bayesian forecasting, both  the support of driving mode management decisions and the issuing of early warnings to the driver are addressed. A statistical modeling framework is created and a suite of algorithms are developed to manage driving modes and issue relevant warnings in accordance with the management by exception principle. The efficacy of these developed methods are then illustrated and examined via a simulated case study.
\end{abstract}




%


\section{Introduction}

Automated driving systems (ADS) are poised to constitute a major technological innovation reshaping transportation as we know it. Recent breakthroughs in Artificial Intelligence (AI), coupled with advances in computational hardware, have had a
revolutionary effect on ADS allowing cutting-edge control algorithms to be executed in real time. 
Despite these advances, it is widely accepted that ADS technology will not be widely deployed in the immediate future; its incorporation onto global roadways will be a gradual but groundbreaking process \citep{Hillier2014, Mahmassani2016}. 

Given these projections, it is important to recall the six-level SAE driving automation taxonomy \citep{SAE2018}. In this taxonomy, level 0 describes vehicles with no automated capacity, and the remaining levels describe vehicles with increasing automated features that culminate in fully automated ADS at level 5. In these level-5 vehicles, the ADS retains control under all road conditions and no design restrictions.
Conversely, in level-3 and -4 vehicles, the ADS may ask the driver 
to take control via a {\em Request-to-Intervene} 
(RtI) command when operational
domain design (ODD) \citep{czarnecki} conditions are not met.
Several estimates indicate that ADSs of these levels will occupy 25\% of the global market by 2040, implying that traffic will be a heterogeneous mix of manual 
and automated vehicles in the coming decades. Until level-5 vehicles  
predominantly populate global roadways, RtI decisions
and their management will remain a crucial, safety-related issue;
see \cite{Caballero} for an overview
of decision support issues related to ADS technology.
 
To ensure the safe operation of level-3 and -4 ADS in the near future, this paper proposes an integrated model for RtIs  under various scenarios. The temporal nature of the RtI decision, in conjunction with the repeated sampling of the environment via ADS sensors, naturally lends itself to decision analysis \citep{frenchrios} and Bayesian forecasting \citep{westharrison}. In this research, these methodological constructs are foundational in determining when to request a transfer of control and help define the degree of decision support required during any intervention. Their utilization also allows for an explicit representation of state uncertainty to reduce the effect of error propagation in ADS operations and decision support.

We begin the development of our driving-mode management approach in Section \ref{secBasic} by defining probabilistic models for the core components of such decisions: the operational design domain, environment and 
driver state monitoring; driving mode assessment; trajectory planning; and
driver intervention performance assessment. These models enable the establishment of associated warnings used to raise driver awareness in subsequent algorithms. In Section \ref{secMgmt}, we develop a suite of 
decision support algorithms to transition between various driving modes based on our developed probabilistic models. In Section \ref{secSim}, we set forth a simulation experiment to examine the efficacy of our approach; the results also highlight several decision-support dilemmas arising in this domain. Section \ref{secDiscussion} provides a discussion of our results and explores avenues of future inquiry.

\section{Core components} \label{secBasic}
Consider a level-3 or -4 ADS for which several 
driving modes are 
available. Example modes include {\em autonomous} (the ADS drives itself), {\em manual} (a person operates the vehicle), or {\em emergency} (the vehicle stops immediately
for safety).  The ADS communicates with the driver 
through a Human Machine Interface (HMI). Via a variety of sensors, the ADS also captures data from the driver and the environment that is utilized 
for ADS management (e.g., driving mode selection).
In this section, statistical models for core ADS management functions are presented including 
ODD monitoring; environment 
monitoring (EM); driver state monitoring (DSM); trajectory planning;
driving mode
assessment (DMA); and, finally, driver intervention performance 
assessment (DIPA).

Accounting for the dynamic features of the collective system is crucial for safe ADS operations.
Predicting departures from the expected or required status quo allows for a more proactive and cautious approach to ADS management. Therefore, utilizing the statistical models developed in this section, we prescribe probabilistic thresholds with associated computational definitions for undesirable or impermissible system states. These characterizations allow for the definition of relevant early warnings that effectively account for the existent uncertainty. Moreover, the 
models underlying the system's dynamics further facilitate ADS management in DMA and DIPA by enabling the use of the decision analysis techniques described herein.

To increase predictive accuracy, a Bayesian approach is utilized based upon observed system behavior. $D_t$ will designate the data set available up to time $t$ incorporated from 
the ADS performance evolution and its sensors. Typically, most ADS maneuvers, tasks and  local trajectory planning are scheduled a few steps ahead, e.g.,  $ k = 10$ time intervals  of $0.5$ seconds. The number  
and length of intervals depend on traffic and driver conditions as well as on the  algorithms' computational demands. As such, these parameters may vary over time; however, their values should, at a minimum, cover the driver's reaction time plus some safety buffer. In our discussion, we assume, without loss of generality, that $k$ is fixed.  

Finally, we note that the notation utilized in this paper is  characterized by the use of bold font for multivariate variables. Unless otherwise stated, upper-case is reserved for random variables whereas lower-case is used for their realizations. Likewise, we reserve $(\cdot)$ for vectors and $[\cdot]$ for matrices. In addition, when convenient, we denote $(y_1, y_2, \dots, y_t)$ as $y_{1:t}$.

\subsection{Operational design domain monitoring} \label{secODD}

An ADS must operate within its
ODD \citep{czarnecki} which constrains the conditions
under which it is designed to function.
These typically include references to
the road environment state (e.g., type of road, presence of temporary
structures, traffic volumes, weather, time,
or visibility conditions);
the behaviour of the ADS (say, speed limits or limitations on maneuvers);
or the vehicle's state (e.g., minimum tire inflation level). 
 Assume that these are checked at time $t$ through
 three blocks
of variables respectively designated 
$\bm{g}_1^t$, $\bm{g}_2^t$ and $\bm{g}_3^t$.
These blocks of variables are concatenated into a single vector such that $\bm{g}_t =(\bm{g}_1^t ,\bm{g}_2^t ,\bm{g}_3^t )\in {\cal G}$ wherein ${\cal G}$ represents the ODD for which the ADS is capable of operating 
autonomously.  

Typically, the evolution of some of the variables can be accommodated with deterministic
physical models. For example, given current position, speed and acceleration,
we could forecast the location and, consequently, whether we will stay within road limits over the next $k$ periods.  However, to maintain generality, we shall use probabilistic models to forecast such variables. 

A flexible strategy for this 
uses state space models \citep{westharrison} through
\[ \bm{G}_t = \phi_t (\bm{F}_t, \bm{V}_t ), \,\,\,\,\,  \bm{F}_t = \gamma_t (\bm{F}_{t-1}, \bm{W}_t ) , \]
where
$\bm{G}_t$ are 
observable
ODD variables (with realisations $\bm{g}_t$); 
$\bm{F}_t$ are unobservable state variables; $\bm{V}_t$
and $\bm{W}_t$ are vectors of random noises; and, $\phi_t$ and $\gamma_t$ are
observation and state functions, respectively. 
Given the short time intervals considered herein,
 an important example is the local
level (i.e., random walk plus noise) model 
\[ \bm{G}_t = \bm{F}_t + \bm{V}_t, \,\,\,\,\, \bm{F}_t = \bm{F}_{t-1}  + \bm{W}_t .  \]
With this information, we build the  forecasting model
$p(\bm{G}_{t+k}|D_t)$ for the ODD variables $k$ periods ahead 
recursively through
\[p(\bm{F}_{t+i}|D_t)=\int p(\bm{F}_{t+i} | \bm{F}_{t+i-1} ) p(\bm{F}_{t+i-1}|D_t)
d \bm{F}_{t+i-1},  \]
for the state variables, and
\[p(\bm{G}_{t+i}|D_t)=\int p(\bm{G}_{t+i} | \bm{F}_{t+i}) p(\bm{F}_{t+i}|D_t)
d \bm{F} _{t+i}, \]
for the ODD variables, with  $i=1,2,...,(k-1), k$. \\


\noindent{\bf ODD-monitoring warnings}. 
Based on the forecasting model $p(\bm{G}_{t+k}|D_t)$,
queries concerning whether $Pr ( \bm{G}_{t+k}\notin {\cal G}| D_t )$ is great enough may be answered sufficiently in advance. In cases wherein the vehicle will likely exceed ODD limits, the ADS should issue an alert and the autonomous mode 
should be abandoned. For the ODD monitoring system, and with subsequent functions discussed later,
we introduce two levels of alerts (i.e., critical and standard warnings) with probabilistic thresholds $ crit _{odd}>
 warn_{odd}$ to improve driver awareness. This implies that
if $Pr ( \bm{G}_{t+k}\notin {\cal G}| D_t )> crit_{odd}$ the ADS issues a
critical warning; otherwise, if
$Pr ( \bm{G}_{t+k}\notin {\cal G}| D_t )> warn_{odd}$, a standard warning is issued.

 A warning can also be issued when an unexpected change in the  operational conditions is detected. For example, if $Pr ( | \bm{G}_{t+1}| > | \bm{g}_{t+1} | ~| ~D_t )$ is sufficiently small, wherein $|\cdot|$ is an appropriate norm,  the ADS may alert the driver of an unexpected change in ODD compliance since the observed conditions were assessed unlikely. 

 
\subsection{Environment monitoring} \label{secEM}
There are also $p$ exogenous, environmental 
conditions $ \bm{Y}_t = ( Y^1_t, \dots, Y^p_t  )$ of interest in a driving scene; some of them may coincide
with the $\bm{g}_1^t$ ODD variables.
They capture properties of the environment that are relevant
for ADS operation. 
Such conditions may reflect, for example, 
the presence of objects, persons, animals or other vehicles in the 
driving scene that are captured and processed through the ADS' sensors and algorithms, respectively.
 The environment variables encapsulate the semantic (i.e., the driving scene identification) and prediction 
(i.e., the driving scene's evolution) layers
in \cite{gal1}. These variables refer to 
highly dynamic conditions that need to be predicted.
More stable environmental conditions (e.g., light and weather) are typically only included within the
ODD variables, specifically $\bm{g}^t_1$ .

We consider a model $ p(\bm{Y}_{t+1}|\bm{Y}_{t}) $ 
which describes  the predictive evolution of the environmental variables between two consecutive time steps.
From it we derive the 
$ k $-steps ahead  
forecast for these variables  through
the recursion 
\[ p(\bm{Y}_{t+i}|D_t) = 
    \int p(\bm{Y}_{t+i}|\bm{Y}_{t+i-1}) 
    p(\bm{Y}_{t+i-1}|D_t)
       d\bm{Y}_{t+i-1}, \quad i=1, 2,..., k-1,k ,\]
summarised by
\begin{equation*}
    p(\bm{Y}_{t+k}|D_t) = p(\bm{Y}_{t+k}|\bm{y}_t) =
    \idotsint \left[\prod_{i=1}^{k} p(\bm{Y}_{t+i}|\bm{Y}_{t+i-1})\right]
    d\bm{Y}_{t+1} \dots  d\bm{Y}_{t+k-1}.  
    \label{eq:Dyn_Forecast_Cov_Ksteps}
\end{equation*}

\noindent{\bf EM warnings}.
Based on the forecasting model $p(\bm{Y}_{t+k}|D_t)$,
once we have observed $\bm{y}_{t+k}$ we may compute whether 
$Pr( |\bm{Y}_{t+k}|>|\bm{y}_{t+k}| \,\, | \,\, D_t)$ is sufficiently small
for a given  
norm $|\cdot|\,$.
If this is the case, the system should warn that the observation evolved in an unexpected manner, suggesting a sudden change 
in the environment. Such alerts are provided to increase driver awareness in potentially hazardous situations. As before,
we assume, without loss of generality, two warning levels triggered by distinct probability thresholds.

\subsection{Driver state monitoring} \label{secDSM}

DSM is an essential component of RtI decisions: 
the ADS must understand the driver state to predict how he/she would react
to an RtI.
For example, a driver who is sufficiently distracted for an extended period of time, may be unable to effectively respond to an emergency situation. Much like the monitoring of the external environment, a 
DSM system includes a variety of sensors used to perceive the driver's state with respect to fatigue and distraction. 

DSM issues have been tackled by authors 
like \cite{Dong2011, Hecht2018} and \cite{Akai2019}. 
Vehicle-oriented approaches use vehicle movements to infer the
driver's state (e.g., acceleration, driving path), whereas human-oriented
methods infer the driver's state 
via his/her actions within the vehicle (e.g., eye closure, gaze direction, hand
position). 
Regardless of the approach adopted, given the substantial effect of driver behavior on roadway safety \citep{Wang2020} 
machine learning models have been a primary focus of recent 
DSM research \citep{Torres2019, Yi2019}, with a few notable examples adopting a Bayesian perspective \citep{Agamennoni2011, Straub2014}. 


We build upon the above by constructing a Bayesian model to monitor the driver's state, issue warnings when this state reaches dangerous levels, and provide predictions for the assessment of driving performance. In principle, the driver's state could be modeled quantitatively or qualitatively using univariate or multivariate variables. Herein, we develop our forecasting framework assuming multivariate,  quantitative driver states, but we also explicitly address warning structures for univariate, qualitative states given their recent emphasis in government and academia \citep{Stutts2001, ranney2001nhtsa, Dong2011}.  

\begin{figure}[htb]
	\centering
	\begin{tikzpicture}[%
	node distance=1.2cm,
	every node/.style={scale=0.5},
	minimum size=1.2cm
	]%
	\tikzstyle{chance}=[circle,draw]
	\tikzstyle{suite}=[->,>=stealth',thick]
	
    \node[] (dots_Ymin1) at (-3,2.5) {$ \dots $};
	\node[] (dots_thetamin1) at (-3,0) {$ \dots $};
	\node[] (dots_Xmin1) at (-3,-2.5) {$ \dots $};

	\node[chance] (Y_t) at (0,2.5) {$ \bm{Y}_t $};
	\node[chance] (theta_t) at (0,0) {$ \bm{\theta}^t $};
	\node[chance] (X_t) at (0,-2.5) {$ \bm{X}_t $};
	
	\draw[suite] (dots_Ymin1) -- (Y_t);
	\draw[suite] (dots_thetamin1) -- (theta_t);
	\draw[suite] (dots_Xmin1) -- (X_t);

	\draw[suite] (Y_t) -- (theta_t);
	\draw[suite] (theta_t) -- (X_t);
	
	\node[chance] (Y_tPlus1) at (3,2.5) {$ \bm{Y}_{t+1} $};
	\node[chance] (theta_tPlus1) at (3,0) {$ \bm{\theta}^{t+1} $};
	\node[chance] (X_tPlus1) at (3,-2.5) {$ \bm{X}_{t+1} $};
	
	\draw[suite] (Y_tPlus1) -- (theta_tPlus1);
	\draw[suite] (theta_tPlus1) -- (X_tPlus1);

	\draw[suite] (Y_t) -- (Y_tPlus1);
	\draw[suite] (theta_t) -- (theta_tPlus1);
	\draw[suite] (X_t) -- (X_tPlus1);
	
    \node[] (dots_Y) at (6,2.5) {$ \dots $};
	\node[] (dots_theta) at (6,0) {$ \dots $};
	\node[] (dots_X) at (6,-2.5) {$ \dots $};
	
	\draw[suite] (Y_tPlus1) -- (dots_Y);
	\draw[suite] (theta_tPlus1) -- (dots_theta);
	\draw[suite] (X_tPlus1) -- (dots_X);

	\end{tikzpicture}
	\caption{Forecasting environment and state}
	\label{fig:Dynamic_Covariates}
\end{figure}
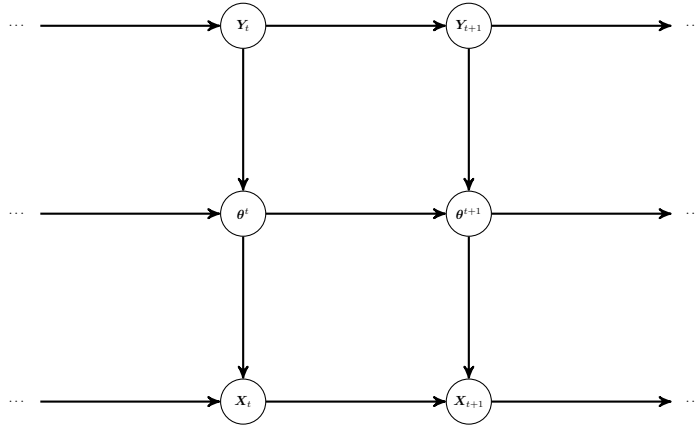

Let the true driver state be a latent variable $ \bm{\theta}^t \in \Theta $. At each time step $t$, the ADS collects $ n $ variables $ \bm{X}_t =(X^1_t, \dots, X^n_t)$ from the driver through sensors
which serve to monitor his/her behavior.
Some of these variables may be continuous (e.g., head position), others may be discrete
(e.g., number of eye blinks).
The environment monitoring variables also affect  
the driver state (e.g., the sudden appearance of an obstacle) and, therefore, should be 
taken into account for monitoring purposes.
Figure \ref{fig:Dynamic_Covariates} depicts the structure of 
our DSM dynamic model.  
In addition to $ p(\bm{Y}_{t+1}|\bm{Y}_{t}) $, 
 the model is specified through
\begin{itemize}
    \item $ p(\bm{\theta}^{t+1}| \bm{\theta}^{t}, \bm{Y}_{t+1}) $,  the predictive evolution of the driver state given her previous 
    state and the exogenous covariates,
    \item $ p(\bm{X}_{t+1}|\bm{\theta}^{t+1},\bm{X}_t) $ which stands for the predictive evolution of the monitoring variables, given the monitoring variable at time $t$ and the future state at time $t+1$, and
    \item $ p(\bm{Y}_1)$, $p(\bm{\theta}^1|\bm{Y}_1)$, and $p(\bm{X}_1|\bm{\theta}^1) $, which are, respectively, the prior 
    over the exogenous covariates, the dependence of the driver state given the initial covariates, and the dependence of the monitoring variable with respect to the initial state.  
\end{itemize}

Under this configuration, whenever period $ t $ starts, we have available 
distributions $ p(\bm{\theta}^{t}|D_t) $ and $ p(\bm{Y}_{t+1}|D_t) = p(\bm{Y}_{t+1}|\bm{y}_{t}) $. The driver state forecast $ \bm{\theta}^{t+1} $ is based on 
\begin{equation}
     p(\bm{\theta}^{t+1}|\bm{Y}_{t+1}, D_{t}) = \int   p(\bm{\theta}^{t+1}|\bm{\theta}^{t}, \bm{Y}_{t+1}) 
      p(\bm{\theta}^{t}|D_t) d\bm{\theta}^{t} ,
    \label{eq:state1}
\end{equation}
\begin{equation}
     p(\bm{\theta}^{t+1}|D_{t}) = \int   p(\bm{\theta}^{t+1}|\bm{Y}_{t+1}, D_t ) 
     p( \bm{Y}_{t+1}|\bm{y}_{t}) d \bm{Y}_{t+1}. 
    \label{eq:state2}
\end{equation}
Moreover, the forecast on the monitoring variable is given by
\begin{equation*}
    p(\bm{X}_{t+1}|D_{t}) = \int  p(\bm{X}_{t+1}|\bm{\theta}^{t+1},\bm{X}_{t})  p(\bm{\theta}^{t+1}|D_t) d\bm{\theta}^{t+1}.
\end{equation*}

At time $ t+1 $, $ \bm{X}_{t+1} $ and $ \bm{Y}_{t+1} $ are observed.
The information update is  $ D_{t+1} = D_t \cup \{ \bm{x}_{t+1},\bm{y}_{t+1} \} $.
Utilizing this information, we next update according to 
\begin{equation}
    p(\bm{\theta}^{t+1}|D_{t+1}) \propto 
    p(\bm{\theta}^{t+1}|D_{t}, \bm{y}_{t+1} ) p(\bm{x}_{t+1}|\bm{\theta}^{t+1},  \bm{x}_{t} ),
    \label{eq:state3}
\end{equation}
    
    
\noindent and the recursion continues.

As mentioned, the ADS will typically 
use a  length $k$ horizon  forecast. 
The driver state forecast $k$ steps into the future is
\begin{equation*}
    p(\bm{\theta}^{t+k}|D_t) = \idotsint  p(\bm{\theta}^{t}|D_{t}) 
    \prod_{i=1}^{k}
    p(\bm{\theta}^{t+i}|\bm{\theta}^{t+i-1},\bm{Y}_{t+i}) p(\bm{Y}_{t+i}|\bm{Y}_{t})
    d\bm{\theta}^{t+i-1} d\bm{Y}_{t+i}.
    \label{eq:Dyn_Forecast_DS_Cov_Ksteps}
\end{equation*}
Similarly, we assess the complete driver state forecast trajectory starting at time $ t+1 $ through
\begin{equation}
 p(\bm{\theta}^{t+1},\dots, \bm{\theta}^{t+k}|D_t) = \idotsint  p(\bm{\theta}^{t}|D_t)
 \left[\prod_{i=1}^{k} p(\bm{\theta}^{t+i}|\bm{\theta}^{t+i-1},\bm{Y}_{t+i}) p(\bm{Y}_{t+i}|\bm{Y}_{t+i-1})
     d\bm{Y}_{t+i} \right]d\bm{\theta}^{t}.
\end{equation}
The above summarises the recursion
(using $p (\bm{Y}_{t+i} |  \bm{Y}_{t+i-1}    )$)
\begin{equation}\label{taiwan}
    p(\bm{\theta}^{t+i}|\bm{Y}_{t+i}, D_t) = \int  
    p(\bm{\theta}^{t+i} |\bm{Y}_{t+i}, \bm{\theta}^{t+i-1})
    p (  \bm{\theta}^{t+i-1}   | D_t) d \bm{\theta}^{t+i-1},  \quad i=1,2,...,k, 
\end{equation}
\begin{equation*}
    p(\bm{\theta}^{t+i}|D_t) = \int  
    p(\bm{\theta}^{t+i} |\bm{Y}_{t+i}, D_t )
    p (\bm{Y}_{t+i} |  \bm{Y}_{t+i-1}    ) d\bm{Y}_{t+i}.  
\end{equation*}
For driver behaviour monitoring purposes, it is also
relevant that
\[ p(\bm{X}_{t+i}|D_t) = 
    \int p(\bm{X}_{t+i}| \bm{\theta}^{t+i},
    \bm{X}_{t+i-1}) 
    p(\bm{X}_{t+i-1}|D_t)
       d\bm{X}_{t+i-1}, i=1,...,k.\]



 
\noindent{\bf DSM warnings.}
   If $\Theta$ is a finite state space with elements $\theta_i$, we write $ \Theta = \{ \theta_1, \dots, \theta_l, \theta_{l+1}, \dots, \theta_m \}$ with values $  \{ \theta_1, \dots, \theta_l \} $ corresponding to \emph{good} states 
  (e.g., aware) and $ \{ \theta_{l+1}, \dots, \theta_m \} $ referring
to \emph{bad}  ones (e.g., sleepy, distracted). 
  When the driver is operating the vehicle or if the operational conditions are near the limit, the vehicle should issue
    a warning if the probability of encountering a bad driver state is great. 
     We can compute whether the predictive probability
of the driver being
in a good state is great enough by considering  a certain threshold $ crit _{dst}$
and checking whether 
$ \sum_{j=1}^{l} p (\theta_{j}^{t+1}|D_{t}) \geq crit_{dst} $.
 A natural value for the threshold is $1/2$, equivalent 
 to checking whether $ \sum_{j=1}^{l} p (\theta^{t+1}_j|D_t ) \geq \sum_{j=l+1}^{m} p (\theta_j^{t+1}| D_t )  $. In this manner, it can be assessed whether it is more likely for the driver to be in a good state than a bad state; conversely, we could be more demanding in relation to the threshold. 
 Similar to other functions, we can implement two alert levels $(warn_{dst}, \
 crit_{dst})$ for issuing warning and critical alarms, respectively.  If  the driver does not respond to these queries and the conditions become too dangerous, the ADS should go into an emergency mode.
 
   Based on the driver behavior monitoring variables, a warning can be issued when an unexpected change is detected (e.g., driver falls asleep). 
   Doing so may be based on whether the probabilities
   associated 
with the monitoring variables fall below a certain level. That is, if
\begin{equation}
Pr (|\bm{X}_{t+1}|> |\bm{x}_{t+1}| \, | \, D_t) \leq \delta ,
\label{eq:Cond_Warning}
\end{equation}
 for a certain $ \delta > 0 $,
with $ |\cdot| $ denoting an appropriate norm,
warnings should be issued based on the unexpected change, possibly to help improve driver awareness. As referenced previously, we consider two warning levels.
 
Furthermore, after $T$ time periods, another quantity of interest 
is the most likely sequence of driver states 
$\bm{\theta}^{1:T} := \lbrace \bm{\theta}^1, \bm{\theta}^2, \dots, \bm{\theta}^T \rbrace$ given the sequence of observed
variables $(\bm{y_{1:T}}, \bm{x_{1:T}})$. This quantity can be  computed 
with the Viterbi algorithm \citep{viterbi}.
Let $V_{t,k}$ be the probability of the most
likely driver state sequence $p(\bm{\theta}^{1:t}, \bm{y_{1:t}}, \bm{x_{1:t}})$ responsible for the first $t$ observations $\bm{y_{1:t}}, \bm{x_{1:t}}$ whose final driver state is $\bm{\theta}^t = k$. Note that, with slight abuse of notation, the $k$-value in this context represents the driver's state, not the forecast length. 
It is straightforward to prove the recursion:
\begin{eqnarray*}
V_{1,k} &:=& p(\bm{\theta}^1 =k , \bm{y}_{1}, \bm{x}_{1}) = p(\bm{x}_{1} | \bm{\theta}^1 = k) p(\bm{\theta}^1 = k |\bm{y}_{1} )p(\bm{y}_{1}) \\
V_{t,k} &=& \max_l p(\bm{x}_{t} | \bm{\theta}^t = k, \bm{x}_{t-1}) p(\bm{\theta}^t = k | \bm{\theta}^{t-1} = l, \bm{y}_{t}) p(\bm{y}_{t} | \bm{y}_{t-1}) V_{t-1,l}.
\end{eqnarray*}
We can sequentially compute $V_{t,k}$,
storing the state $\pi(k,t)$ used to compute the last equation, for each $k$. Then, the most likely sequence of driver's states $\bm{\theta_*}^{1:T}$ is obtained using the recursion
\begin{eqnarray*}
\bm{\theta_*}^T &=& \arg\max_{k} V_{T,k}, \\
\bm{\theta_*}^{t-1} &=& \pi (\bm{\theta_*}^t, t).
\end{eqnarray*}
Once $\bm{\theta_*}^{1:T}$ is computed, 
we can raise a warning/critical alarm
if the number of time periods in which the most likely state
is bad exceeds a predefined threshold.

\subsection{Trajectory  planning} \label{secTraj}

The ADS will have available a trajectory planning system, see \cite{Gonzalez2016, Claussmann2019, Katrakazas2015} for reviews. At a given instant $ t $, and for a length $ k $ horizon forecast, it provides the ADS with a trajectory plan $ \bm{\Bar{z}} = \{ \bm{z}_{t+1}, \bm{z}_{t+2}, \dots, \bm{z}_{t+k} \} $, within the ODD boundaries (e.g., stay within road boundaries, keep safety distances, do not collide, etc.)

\subsection{Driving mode assessment} \label{secDMA}

Relatively few studies focus on the RtI decision. \citet{mccall2019taxonomy} provide a taxonomy of driving mode transitions (e.g., scheduled, driver- or system-initiated emergency) and discuss its relation to the SAE automation taxonomy. Whereas multiple authors have examined the effect of HMIs on RtIs \citep[e.g.,][]{walch2015autonomous, eriksson2017takeover}, the algorithmic specifics related to the management of driving modes are less developed. 

Consider assessing a driving mode $d$
for a length $ k $ horizon at a given instant $ t $.
Assume we have a trajectory plan
$ \bm{\bar{z}} $ .
 The following relevant ingredients are available:
a forecast on the exogenous environmental states $ \bm{\Bar{Y}} = [ {\bm Y}_{t+1}, \dots, {\bm Y}_{t+k} ] $ (Section \ref{secEM}) and a forecast of the driver state $ \bm{\Bar{\theta}} = [ \bm{\theta}^{t+1}, \dots, \bm{\theta}^{t+k} ] $
(Section \ref{secDSM}). Based on them, we define
\[
p(\bm{\Bar{Y}},\bm{\Bar{\theta}}|D_t) = p(\bm{Y}_{t+1} | \bm{Y}_{t})p(\bm{\theta}^{t+1} | \bm{Y}_{t+1}, D_t )
\prod_{i = 2}^{k} p( \bm{\theta}^{t+i} | \bm{\theta}^{t+i-1}, \bm{Y}_{t+i}, D_t) p(\bm{Y}_{t+i} | \bm{Y}_{t+i-1}).
\]
A utility function 
 $u(d,\bm{\bar{z}},\bm{\Bar{Y}},\bm{\Bar{\theta}}, \bm{g}_t)$ 
 assesses the efficacy of
 the driving mode over the next $k$ steps,  
 incorporating the objectives and 
information for times $ t+1 $ to $ t+k $.
It thus takes into account
 the driving mode, trajectory, environment
 forecast, driver state forecast, and last observed ODD conditions. Typical consequences assessed would include 
 comfort, internal and external safety, 
fuel consumption, travel time and reaction time.
The importance of various objectives would 
usually change depending on the
driving environment. For example, higher speeds 
typically provide more utility on a highway than in
a residential area.

Given the uncertainty, we compute the expected utility of the driving mode 
through 
\begin{equation}\label{marrakech}
\psi(d) = \int \int 
u(d,\bm{\bar{z}},\bm{\Bar{Y}},\bm{\Bar{\theta}},\bm{g}_t)
\ p(\bm{\Bar{Y}},\bm{\Bar{\theta}}|D_t) \ d\bm{\Bar{Y}} \ d\bm{\Bar{\theta}}, 
\end{equation} 
which constitutes the basic ingredient for driving mode assessment so that a mode yielding the greatest expected utility is preferable. Note that, given the gradual evolution of ODD conditions and the compressed nature of the planning horizon, it is not necessary to forecast the ODD variables for the assessment decision.

Since this is the most computationally expensive calculation to be performed, we may introduce a model to forecast driving mode assessments. Such forecasts, call them 
$\widehat{\psi} (d)$, can consequently speed up the process of selecting the preferred driving mode. Of course, warnings in relation to $\bm{Y}_t$ associated with important changes in the environment would redirect attention to the full assessments $\psi$ rather than their forecasts $\widehat{\psi}$.

\subsection{Driver intervention performance assessment}

Additional system inputs beyond sensors with myopic perception have been proposed. Specifically, \textit{driver intervention performance assessment} (DIPA) systems are used to record and retain historical data regarding driver interventions. This information can be used to inform future RtI decisions \citep{Bianchi2019}. 

To wit, after an RtI is issued, ADS operations can be improved if the driver's performance can be evaluated.
For a given $ t $ over the interval $ [t+1, t+k] $, the utility associated with the actual performance can be computed. Let this value be  $u(d_1)$ wherein $d_1$ designates the manual driving mode. This quantity represents the {\em driving score}. After the ADS transfers control to the driver, $u(d_1)$ is compared to their expected performance, $\psi(d_1)$, to asses the intervention. We consider both discrete and continuous methods of DIPA modeling under this framework.

Our approach to DIPA is characterized by identifying whether a driver under- or overperformed with respect to their expected performance. This is accomplished by examining the difference between  $u(d_1)$ and  $\psi (d_1)$ as follows:
\[
    \begin{cases}
      \psi(d_1) - u(d_1) > 0, & \text{ underperformed,} \\
      \psi(d_1) - u(d_1) \leq 0, & \text{ overperformed.}
    \end{cases}
\]

In the discrete approach, we draw inferences about the probability $q$ of the driver underperforming. 
A simple version uses the beta-binomial model \citep{frenchrios}:
if there were $h$ underperformances out of $n$ RtIs such that $D_t =\{ h,n \}$, the posterior is
$q| D_t \sim \beta eta (d + h, e + (n-h))$
wherein $d$ and $e$ are prior parameters.
We may also include $q$ in the  utility function, or an estimate ${\hat{q}}$ (e.g., its  mean), to make it more risk averse as $q$ increases. For example, we could use $u^{1/q}$ wherein $u$ is the original utility function (assuming without loss of generality that $u\in [0,1]$). 

In the continuous case, the focus is on the utility differences: $ \zeta= \psi(d_1) - u(d_1)  $. We can draw inferences on the mean utility difference $\mu $ based on a normal-normal model \citep{frenchrios}. 
Indeed, if there were $n$ RtI cases
with observed performances $\zeta_1, \ \zeta_2,...,\ \zeta_n$, we can assume a prior $ \mu \sim \mathcal{N}(\mu _0 ,\tau^2_0 ) $ and a likelihood function $\zeta_i|\mu, \sigma^2  ~\sim~\mathcal{N}\left(\mu,\sigma^2 \right) $,
resulting in a posterior $ \mu|D_t ~\sim~\mathcal{N}
\left( \mu_1= \frac{\sigma^2 \mu _0 + n\tau^2_0  \Bar{\zeta}}{n\tau^2 _0 + \sigma^2}, \frac{\sigma^2 \tau^2_0 }{n\tau^2_0 
+ \sigma^2}\right) $ wherein $ \Bar{\zeta} $ is the performance sample
mean. 
Subsequently, some factor could be included in the utility function to increase risk aversion when the
driver is expected to underperform. 
For example, if we use the posterior mean $\mu_1 $,
we could retain the original utility function $u$ if $\mu_1 \leq 0$ (i.e., an overperformance is expected) and use $u^{\frac{1}{1+\mu_1}}$, otherwise (assuming, again, that the utility is scaled within [0,1]). Such a definition ensures the utility function becomes increasingly risk averse as the expected underperformance worsens.

\noindent{\bf DIPA warnings.} Consider now warnings related to DIPA. In the discrete approach, if the value of $q$ is great enough, e.g., as checked through a condition of the form
$ Pr ( q  \geq \beta  | D_t  ) > \alpha $, 
for $\alpha$ and $\beta $ sufficiently large, the ADS may advise the person that their driving skills are insufficient for the present scenario and that additional training is advisable. This information could also be used in driving mode management decisions. 

In the continuous approach, if the posterior gives more probability to underperformance,
computed through $Pr ( \mu > 0 | D_t) > 0.5$, the ADS may advise the person that their driving skills are insufficient for the present scenario and that additional training is advisable.



\section{The management of transitions between driving modes} \label{secMgmt}

\begin{figure}[htbp!]
    \centering
    \includegraphics[width=0.8\textwidth]{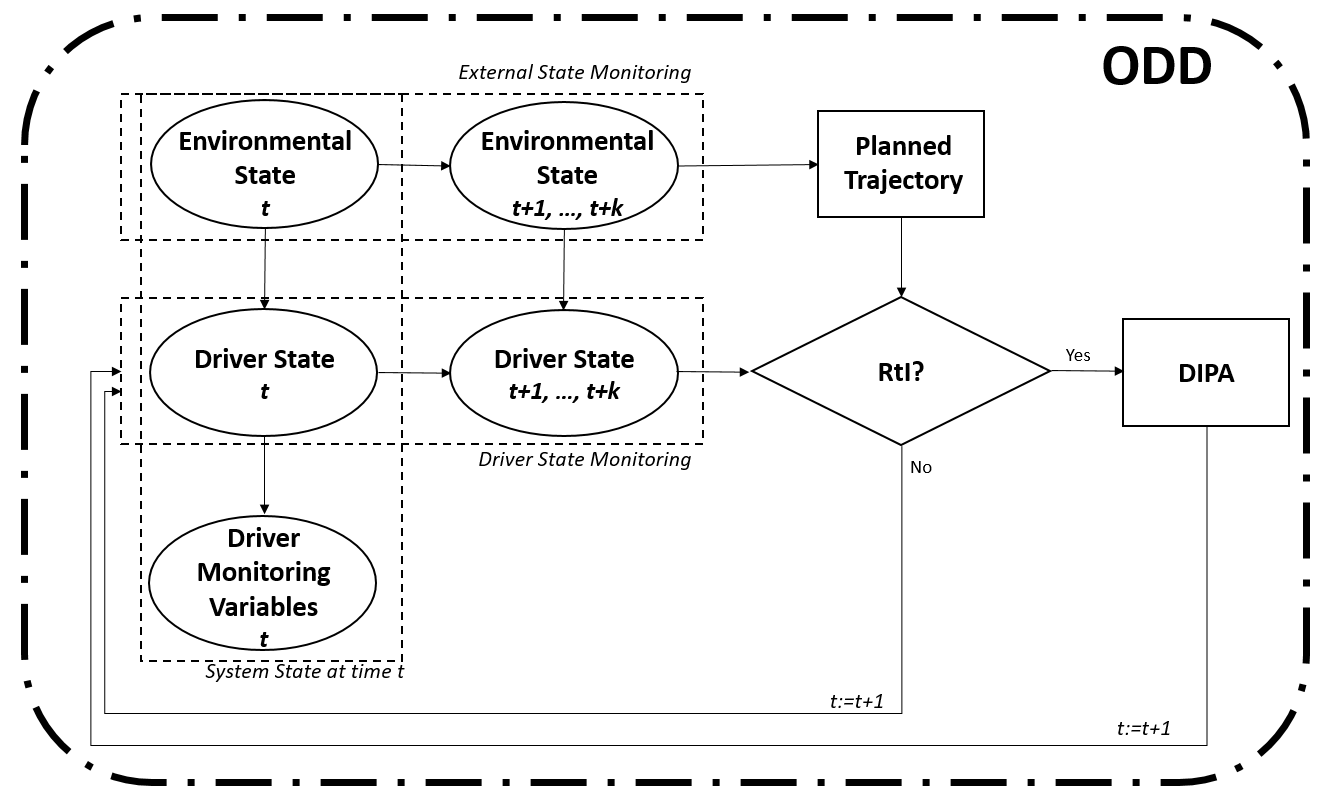}
    \caption{Incorporation of DIPAs into the RtI Decision Process}
    \label{figDIPAinRti}
\end{figure}

Building on the previously derived forecasts, we present a decision-analytic framework that tackles the RtI issue and the management of driving modes. Figure \ref{figDIPAinRti} illustrates the process utilized herein to incorporate decision support based on the introduced elements. The environmental and DSM systems observe the state at time $t$, forecast future states through time $t+k$, update the planned trajectory, and decide whether an RtI should be executed and assessed via the DIPA. 

The approach in Figure \ref{figDIPAinRti} emphasizes a {\em management by exception principle} \citep{westharrison} wherein a group of models is used for inference, prediction and decision support under standard driving circumstances until an exception arises that triggers an intervention request. 
The corresponding warnings can be modulated in two directions: (1) several alert levels (e.g., warning and critical) can be introduced for each relevant driving safety issue; (2) if an alert must be issued repeatedly, the HMI can successively amplify the alert (e.g., increase volume of subsequent warnings).

Under this framework, we assume driving states are univariate and qualitative in accordance with \citet{Stutts2001} and \cite{ranney2001nhtsa}. Likewise, we consider three modes (i.e., autonomous, manual and emergency) respectively designated as $d_0$, $d_1$ and $d_2$, when convenient. Potential transitions between these modes are discussed in this section.

\subsection{Transitions from autonomous} \label{secAuto}
Start with transitions from autonomous 
mode. A planned trajectory, DSM, environment, and vehicle operation inputs
are available. 
The system periodically issues predictive
risk assessments based on compliance with the 
ODD. If $Pr ( \bm{g}_{t+k}\notin {\cal G} | D_t )$ is sufficiently great,
the ADS alerts the driver,
and assesses the autonomous and manual driving modes.
If $\psi (d_0) > \psi (d_1)$, the autonomous mode is preferable; however, as the
ADS is critically approaching its design limits, the
system should enter the emergency mode and issue the appropriate alert.
If not, an RtI is issued to the driver through the HMI followed by a DIPA. If the driver performs too poorly, as assessed from the continuous DIPA model,
the driver is perceived not to be in good condition and the emergency mode
is triggered. Otherwise, the driver takes back control 
until further notice.

\subsection{Transitions from manual} \label{secMan}

In the manual mode, the planned trajectory operates in the background (since the driver is in control), but we have the DSM
and environment inputs in the foreground.
At a certain point, the driver requests the autonomous mode. To ensure safety, non-compliance with operational domain conditions is assessed by checking 
whether $Pr ( \bm{g}_{t+k}\notin {\cal G} | D_t )$ is great enough. If so,
the ADS issues a warning; if ignored by the driver,
it may enter emergency mode. That is, the system checks whether $\psi (d_0) > \psi (d_1)$; if this condition is false, the system
assesses that manual mode is preferable and alerts the driver. If the warning is neglected, the ADS enters 
in the emergency mode.
Conversely, if $\psi (d_0) > \psi (d_1)$, the system enters in the autonomous mode.

Although not considered here, the driver could also directly request a
transition to emergency mode. The system would
 perform computations concerning whether operational conditions
are met with sufficiently high probability and whether 
the autonomous mode
is expected to yield better performance than
manual mode. If deemed appropriate, the ADS could then transition to autonomous. 

\subsection{Transitions from emergency} \label{secEmerg}

The emergency mode is autonomous  
with just  one criteria (i.e., safety) and a computationally fast trajectory planning routine to route the car 
to a safe spot, stopping as fast as possible, while taking into account ethical issues in case of an unavoidable collision. 
 Once stopped, the driver may request to activate the autonomous or manual modes, assuming conditions allow for driving resumption (e.g., the vehicle has not suffered a catastrophic failure, passengers are not injured, etc.).
 The ADS will assess whether the operational conditions are met,
 and whether the driver state is adequate. If driving is feasible, it will evaluate which of the two modes is preferable
 and proceed accordingly. If driving is not feasible, it will warn about this and remain idle.

Of note, the incorporation of ethics when pursuing the emergency mode is a nuanced issue.  \citet{lin2016ethics} presents several vignettes illustrating moral dilemmas associated with ADS, which 
point designers to important problems. For example, in a situation wherein a collision with a pedestrian is unavoidable,  is it more ethical to swerve right,  protecting the  occupants, or left, protecting a greater number of pedestrians? Ethical dilemmas such as these are collectively known as \textit{trolley problems} and are of long-standing interest  \citep[e.g., see][]{Bauman2014, Eyal2020}.
Although the use of such problems has been criticized as lacking relevance to algorithm training \citep{basl2020}, they are widely used as an analogy in the study of ADS ethics. Importantly, \citet{Awad2018} developed the \textit{Moral Machine} platform,
wherein users are shown trolley problems and asked to make 
driving choices. 

\subsection{Summary}

The (mostly qualitative) scheme in Algorithm \ref{alg:controller} summarises the complete ADS management procedure. There are three modes AUTON, MANUAL, EMERG.
Commands on the same line are processed in parallel. The variables $\psi_0$ and $\psi_1$ designate the assessment ($k$ steps ahead) of the AUTON and MANUAL modes, respectively. 

{\tiny 
\begin{algorithm}[!htb] %
\caption{ADS controller}  
\label{alg:controller}
\begin{algorithmic}
\State {\bf Input:} Priors for ODD, environment, driver state.
Utility function 
\State {\bf Output:} Trajectory from ORIGIN to DESTINY
(and implementation of commands when in AUTON or EMERG modes). 
\While{DESTINY not reached}
    \State Read internal sensors. Read external sensors. 
    \State Forecast Environment $k$ steps ahead. Forecast driver state $k$ steps ahead. Compute trajectory.
    \State Assess driving modes ($\psi_0$, AUTON; $\psi_1$, MANUAL). Issue WARNINGS.
    \State Manage from DRIVING MODE.  If any DIPA pending, resolve it
\EndWhile
\end{algorithmic}
\end{algorithm}
}

{\small 
\begin{algorithm}[!h] %
\caption{Issuing WARNINGS}  
\label{alg:warnings}
\begin{algorithmic}
\State {\bf Input:} Thresholds for warnings ($warn_i , crit_i$)
\State {\bf Output:} Warnings to be issued. \\
Compute $q_1=Pr ( \bm{G}_{t+k}\notin {\cal G}| D_t )$; $q_2=Pr ( | \bm{G}_{t+1}| > |\bm{g}_{t+1}| | D_t )$, $q_3=P( |\bm{Y}_{t+k}|> |\bm{y}_{t+k}| | D_t)$,
$q_4= \sum_{j=1}^{l} p (\theta_{j}^{t+1}|D_{t})$, $q_5= Pr (|\bm{X}_{t+1}|> |\bm{x}_{t+1}| |D_t)$.

\If{$q_1 > crit_{odd}$ }
            \State Warning {\em About to leave ODD limits}
            \ElsIf{   $q_1 > war_{odd}$      }
            \State Warning {\em Reaching ODD limits}
\EndIf
\If{$q_2 < crit_{veh}$ }
            \State Warning {\em Very odd vehicle behaviour}
            \ElsIf{   $q_2 < war_{veh}$      }
            \State Warning {\em Odd vehicle behaviour}
\EndIf
\If{$q_3 < crit_{env}$ }
            \State Warning {\em Very odd environment behaviour}
            \ElsIf{   $q_3 < war_{veh}$      }
            \State Warning {\em Odd environment behaviour}
\EndIf
\If{$q_4 < crit_{dst}$ }
            \State Warning {\em Very bad driver state}
            \ElsIf{   $q_4 < war_{dst}$      }
            \State Warning {\em Bad driver state}
\EndIf
\If{$q_5 < crit_{dbh}$ }
            \State Warning {\em Very odd driver behaviour}
            \ElsIf{   $q_5 < war_{dbh}$      }
            \State Warning {\em Odd driver behaviour}
\EndIf
\end{algorithmic}
\end{algorithm}
}

Algorithm \ref{alg:warnings} manages the issue of warnings. Its five conditional statements may be processed in parallel but are presented sequentially for clarity of exposition. The issuing of warnings coincide with the discussions in Section \ref{secBasic}. Algorithms \ref{alg:emerg}, \ref{alg:auton} and \ref{alg:manual} formalize the management of driving modes as discussed Sections \ref{secEmerg}, \ref{secAuto} and \ref{secMan}. Notably, once the ADS enters EMERG mode, it remains there until the vehicle stops. Otherwise, the ADS will implement the decisions in AUTON or MANUAL mode. Finally, Algorithm \ref{alg:dipa} sketches the procedure of executing a DIPA and issuing the associated alerts.

\begin{algorithm}[!h] %
\caption{Managing from EMERG}  
\label{alg:emerg}
\begin{algorithmic}
\State Compute ethical trajectory
\State Implement trajectory
\State STOP
\end{algorithmic}
\end{algorithm}
\begin{algorithm}[!h] %
\caption{Managing from AUTON}  
\label{alg:auton}
\begin{algorithmic}
\State {\bf Input:} $q_1$, $\psi_0$, $\psi_1$. 
\State {\bf Output:} Mode to be followed
\If{$q_1>crit_{odd}$}
        \If{$\psi _0 > \psi_ 1 $}
        \State Undertake EMERG
        \Else 
        \State Issue RtI, Start DIPA
        \EndIf
   \Else
    \State Remain AUTON
\EndIf
\end{algorithmic}
\end{algorithm}
\begin{algorithm}[!h] %
\caption{Managing from MANUAL}  
\label{alg:manual}
\begin{algorithmic}
\State {\bf Input:} $q_1$; $\psi_0$; $\psi_1$. 
\State {\bf Output:} Mode to be followed
\If{Driver requests AUTON}
    \If{$q_1>crit_{odd}$}
        \If{$\psi _0 > \psi_1 $}
            \State Start EMERG
            \Else
            \State Warning {\em Better MANUAL}
            \If{Warning neglected}
                \State Start EMERG
                \Else
                \State  Remain MANUAL
            \EndIf
        \EndIf
        \Else
        \State Start AUTON
    \EndIf 
    \Else
    \State Remain MANUAL
\EndIf
\end{algorithmic}
\end{algorithm}

\small{
\begin{algorithm}[!h] %
\caption{Resolving a DIPA}  
\label{alg:dipa}
\begin{algorithmic}
\State {\bf Input:} $\psi_1$, d, e.
\State {\bf Output:} Assess $u(d_1)$. 
\If{$\psi_1 >u (d_1) $}
                \State $d=d  +1$. 
        \Else 
        \State $e = e+1$. 
        \EndIf
       \State Compute $q_6 = Pr (\beta eta (d,e) \geq \beta )$
 \If{$q_6 > crit_{dipa}$ }
            \State Warning {\em Very bad driving performance}
            \ElsIf{   $q_6 > war_{dipa}$      }
            \State Warning {\em Bad driving performance}
\EndIf      
 \end{algorithmic}
\end{algorithm}
}

\section{Simulation} \label{secSim}
This section specifies a simulation environment used to
globally assess the proposed ADS control mechanism.
We describe first how the driving environment is configured and 
then how the controller is implemented. Although the
logic remains the same, simplified versions of the core components
are provided which would be replaced by actual components in a real environment.

This section also serves to illustrate what can be referred to as a \textit{fundamental dilemma} in level-3 and -4 ADS. If a risky situation is predicted, the driver is expected to be aware; however, if the driver is actually
distracted, consequences could be catastrophic. When considering such situations during ADS design, it is unclear whether it is preferable to transfer control to the driver allowing the human to assume the risk associated with their distraction, or should a conservative approach be adopted via the continuation of AUTON mode. Moreover, in the latter case, would the driver or the car manufacturer be held liable in case of an accident? Whereas the question of liability is likely a philosophical matter, we illustrate how the previous two questions can be addressed via expected utility maximization.

\subsection{Driving Environment Configuration}

The driving environment configuration describes the road and vehicle attributes, probability distributions over the environment's dynamics, the support upon which the forecasts are constructed, as well as trajectory and utility assumptions needed to determine RtI decisions. 
\paragraph{Road configuration.}
    The ADS moves along a single lane straight road  decomposed into cells of equal length.
\paragraph{Vehicle configuration.}
The ADS is level-3 with two modes AUTON and MANUAL.
    In AUTON mode, it is able to select a speed. It has four speeds 0 (stop), 1, 2, 3 (number of cells advanced per time interval).  Speed choices are transmitted immediately at the beginning of the time interval.
    
\paragraph{Driver features}
    The driver (she) may be in two states:
    $\theta _1$ (distracted) and $\theta _2$ (aware). 
    In MANUAL mode, the driver is also able to select a 
    speed, with five possible speeds:
    0 (stop), 1, 2, 3, 4 (number of cells     advanced per time interval).  Speed choices are 
    transmitted immediately at the beginning of the time interval if     the driver is aware and one time interval after if she is distracted. When aware, the driver requires one time interval to respond to an RtI; she requires three time intervals if distracted.

\paragraph{Environment (Section \ref{secEM}). }
 Two types of objects may appear in the driving scene:
\begin{itemize}
    \item A red object that covers one cell is associated with 
    a rock. The driver must stop to avoid crashing with it. Once stopped, the rock disappears and the vehicle may restart.
    \item A blue object that also covers one cell is associated with a puddle. If the car crosses it at maximum speed in  AUTON mode (i.e., speed 3) it will likely skid,
    specifically with probability 0.95. If the driver is distracted and drives at speed 2, 3, or 4 it will skid with probability 0.5, 0.8 or 0.85, respectively.
\end{itemize}
Table \ref{kakaroad} reflects the probabilistic dynamics of the obstacles, $p(Y_{t+1} \, \vert  \, y_t)$. For example, 
after a rock, the road will necessarily be clean. 

    {\small
\begin{table}[h!]
	\centering
	\begin{tabular}{c|c|c|c}
			        &  Rock  & Water  &  Clean  \\ 		\hline  
Rock &  0 & 0     &   1   \\
Water   & 0   &   0.4  &   0.6             \\ 
Clean &  0.05 &  0.05 & 0.90 \\ \hline
		\end{tabular}%
	\caption{Probabilistic evolution of road state. Row current cell; 
	Column next cell}
	\label{kakaroad}%
\end{table}
}

\noindent Assume that objects are perceived without confusion some time in advance as reflected in Table \ref{kaka2}. For example, in the AUTON mode, car sensors detect 
the presence of a rock three cells in advance. Perceptions in MANUAL mode depend on whether the driver is distracted (dist) or aware. 

    {\small
\begin{table}[h!]
	\centering
	\begin{tabular}{c|c|c}
			        &  Rock  & Water    \\ 		\hline  
AUTON &  3 & 1        \\
MANUAL (dist)   & 2  &   Not perceived               \\ 
MANUAL (aware)  &  5 &  5  \\ \hline
		\end{tabular}%
	\caption{Distance detection of objects.}
	\label{kaka2}%
\end{table}
} 
\paragraph{ODD (Section \ref{secODD}).}
 The ODD requires the ADS not to drive at its maximum speed (i.e., speed 3) when in a puddle since would very likely skid. The minimum separation from an obstacle  must be three cells.
\paragraph{Driver state (Section \ref{secDSM}).}
 The ADS senses a discrete random variable $X_t$, say number of blinks per time interval, that informs about the driver state. This monitoring variable has three values. The probability $p(X_t \vert \theta_t)$ of any of them given the driver's state is shown in  Table \ref{kakawink}. Unlike Section \ref{secDSM}, $X_{t}$ does not depend on $X_{t-1}$.
    
    {\small
\begin{table}[htbp]
	\centering
	\begin{tabular}{c|c|c|c}
			        &  $X_t=1$  & $X_t=2$  &  $X_t=3$  \\ 		\hline  
Distracted &  0.1 & 0.2    &   0.7  \\
Aware      	&    0.7  &   0.2  &   0.1             \\  \hline
   	        					\end{tabular}%
	\caption{Probabilities of blinks given the state.}
	\label{kakawink}%
\end{table}
}
We also present the awareness evolution, $p(\theta_{t+1} \vert y_{t+1}, \theta_t )$, in Table \ref{kakaware}. This value depends on the earlier awareness state and current obstacles. For example, if the driver is aware and the next cell is clean, there is a probability 0.15 that she will be distracted in the next period.

    {\small
\begin{table}[h!]
	\centering
	\begin{tabular}{c|c|c|c}
		Current & Obstacle	        &  Aware  & Distracted    \\ 		\hline  
Aware& Rock &            1     &    0           \\
     & Puddle   &       0.99     &  0.01               \\ 
    & Clean    &         0.85   &       0.15           \\ 
Distracted& Rock &       0.95   &    0.05      \\
     & Puddle   &    0.75      &     0.25             \\ 
    & Clean    &       0.05  &      0.95           \\ 
		\end{tabular}%
	\caption{Probabilistic evolution of driver awareness, given current state
	and presence of obstacles.}
	\label{kakaware}%
\end{table}
}
\paragraph{Trajectory planning (Section \ref{secTraj})}
Trajectory planning is performed $k=5$ time intervals in advance
and assimilated with speed choices over the next five time intervals.

\paragraph{Preference model (Section \ref{secDMA})}
The utility function is defined such that its value increases with the ADS' speed. However, the ADS should avoid 
collisions (for safety reasons) and maximize the time in AUTON mode (for comfort reasons).

Table 
\ref{kakaspeed} reflects the utilities 
attained at various speeds per cell. For example, 
a cell crossed at speed 1 provides utility 0.1.
    {\small
\begin{table}[h!]
	\centering
	\begin{tabular}{c|c|c|c|c|c}
			        &  0 & 1 & 2 & 3 & 4  \\ 		\hline  
Speed &  0 & 0.1 & 0.2 & 0.3 & 0.5  \\  \hline
		\end{tabular}%
	\caption{Evaluation of cells.}
	\label{kakaspeed}%
\end{table}
}

\noindent Each collision with a rock is associated with a utility of -100;
each skidding incident incurs a utility of -10. 
Moreover, each cell driven in autonomous mode provides a utility of 0.1.
The assessment
of a trip's section is computed by summing the utilities over
its cells.
\paragraph{Simulating the driving environment}
To simulate the environment, we pick a road size, e.g.,
200 cells. The first cell is initialized as clean. Thereafter, 
we deduce the presence of puddles and rocks based on Table \ref{kakaroad}. Once the driving environment is established,
we simulate the awareness of the driver from Table \ref{kakaware}, assuming the 
driver is initially aware. Subsequently,
we simulate the driver's blinks in accordance with Table \ref{kakawink}.


\subsection{Model implementation} \label{secModelImplement}

Within the above environment, we implement 
a driving decision making controller based 
on the following models.
\paragraph{ODD and environment monitoring (Sections \ref{secODD} and \ref{secEM})}
ODD and environment monitoring are performed as follows. For the next three positions, we sense rocks perfectly; a warning is issued immediately regarding their presence. In contrast, puddles are sensed only one cell in advance. Thus, we must build models to infer the probability of finding puddles in cells $t+3$ and $t+2$ given the content of cells $t+1$. In addition, there is uncertainty about the content of cells $t+4$ and $t+5$.

The elements of Table \ref{kakaroad} are not known by the ADS.
Therefore, producing forecasts for the contents in the next five cells (i.e., accomplishing ODD and environment monitoring) requires learning the environment dynamics. For this, we use Dirichlet-multinomial models \citep{frenchrios}. 
Let us designate $\bm{p}_{i} = p(y_{t+1} | y_t = i)$, where $y_{t+1}$ denotes the obstacle in cell $t+1$ and $i$ takes values 0, 1, 2, for rock, puddle, and clean, respectively.
Uncertainty about $\bm{p}_{i}$
is modelled through a Dirichlet distribution such that $\bm{p}_{i} ~\sim~ \text{Dir}(\bm{\alpha}_i)$. A priori, we set $\bm{\alpha}_i = \bm{1}$. When the ADS moves one cell forward, it updates the current estimate of $\bm{p}_{i}$ in the following manner.
Assume the ADS enters into cell $t$; it knows $y_{t}=i$, the obstacle present in that cell, if any, and immediately observes $y_{t+1}$, the content of cell $t+1$, regardless of whether it is a rock, a puddle, or clean. With this information $\bm{p}_{i}$ is updated to 
$    \bm{p}_{i} \vert y_{1:t+1} \sim \text{Dir}(\bm{\widehat{\alpha}}_i)
$
where $\bm{\widehat{\alpha}}_{i,y_{t+1}} = \bm{\alpha}_{i, y_{t+1}} + 1 $. 
After this update is performed, predictions for cells $t+2$, $t+3$, $t+4$ and $t+5$ can be made in a sequential fashion. This is accomplished using the current estimate of $\bm{p}_{i}$, taking into account whether cells $t+2$ or $t+3$ contain a rock, and making subsequent observations. To simplify computations, let us focus on single-point estimates of $\bm{p}_i$  
approximating $p(y_{t+1} = j | y_t = i)$ by  $\mathbb{E}[\bm{p}_{i,j}] = \frac{\bm{\alpha}_{i, j}}{\sum_k \bm{\alpha}_{i, k} }$. As an example, consider making a prediction for the content of cell $t+4$. We distinguish 
two cases:

\begin{itemize}
\item Suppose cell $t+3$ contains a rock. Since this can be perceived 
through the ADS sensors, the prediction for cell $t+4$ is 
\begin{equation*}
    p(y_{t+4} = j | y_{t+3} = i) = \frac{\bm{\widehat{\alpha}}_{i, j}}{\sum_k \bm{\widehat{\alpha}}_{i, k} }
\end{equation*}
\item Assume cell $t+3$ has a either a puddle or is clean. Since the ADS sensors cannot perceive the contents with certainty, we have 
\begin{equation*}
    p(y_{t+4} = j | y_{1:t+1}) = \sum_{y_{t+3}=0}^2 p(y_{t+4} = j | y_{t+3}) p(y_{t+3} | y_{1:t+1})
\end{equation*}
where $p(y_{t+3} | y_{1:t+1})$ can be computed again considering two cases, depending on the content of cell $t+2$. In turn, this computation will entail calculating $p(y_{t+2} | y_{1:t+1})$, thus reaching the end of the recursion. The quantity $p(y_{t+2} | y_{1:t+1})$ again has two cases: if a rock is observed in cell $t+2$ then $p(y_{t+2} = 0 | y_{1:t+1})= 1$, and 0 otherwise.
If not,
\begin{equation*}
    p(y_{t+2} = j | y_{1:t+1} ) = p(y_{t+2} = j | y_{t+1} = i )  \propto \frac{\bm{\widehat{\alpha}}_{i, j}}{\sum_k \bm{\widehat{\alpha}}_{i, k} }
\end{equation*}
where $i$ is the observed content of cell $t+1$ and $j$ is either a puddle or clean.
\end{itemize}

\paragraph{DSM monitoring (Section \ref{secDSM})}
%
No learning is possible for this monitoring as the driver state is not observable (except in experiments).  Therefore, we assume that the models in Tables 3 and 4 are known by the system. When the ADS moves one cell forward, it will produce an estimate $p(\theta_t \vert D_t)$ of the driver's state given all the information available up to that point (i.e., $x_{1:t}$ and $y_{1:t}$). This can be computed recursively using equations \eqref{eq:state1}, \eqref{eq:state2}, and \eqref{eq:state3}. Additionally, a forecast of the driver's state in cells $t+1$ through $t+5$ is performed using the recursion \eqref{taiwan}.

\paragraph{Trajectory planning (Section \ref{secTraj})}
At each cell, the ADS produces forecasts regarding what obstacles will be present in the next five cells using the aforementioned Dirichlet-multinomial models. In particular, the ADS knows with certainty whether the next cell has a puddle and if any of the next three cells has a rock. We pick the most likely cell configuration and select the speed maximizing expected utility at each cell, thereby providing the trajectory. This plan is updated with each cell the ADS moves forward.

\paragraph{Manual driving planning (Section \ref{secDMA})}
Manual driving planning is performed in a similar fashion except that it is based on the driver's perception and without a forecast:
the driver examines the scene and, if aware, rocks and puddles are perceived up to five cells ahead;
on the other hand, if distracted, rocks are perceived up to two cells ahead and puddles are not perceived. The driver selects the maximum speed by taking into account skid probabilities and costs:
if a rock is sensed, the driver will stop; 
if she encounters a puddle while driving at speeds 0, 1, 2, 3 or 4,  a skid will occur with probabilities 0.0, 0.0, 0.5, 0.8 and 0.85, respectively. These provide expected utilities of 0, 0.1, -4.8, -7-7 and -8.1. Consequently, the driver will choose speed 1. If road is clean, she will choose maximum speed 4. Recall that when the driver is unaware, decisions take one time slice to be transmitted.

\paragraph{ADS Control (Section \ref{secMgmt}).}

Using the previously defined elements, and following the scheme presented in Algorithm \ref{alg:controller}, the trip is managed as follows. (DIPAs are not performed in this example).

\begin{enumerate}[labelindent=*, style=multiline, leftmargin=*]
    \item The ADS enters into the first cell. It has knowledge about the elements of Tables \ref{kakawink} and \ref{kakaware}. The utility function of Table \ref{kakaspeed} is also known. The Dirichlet distributions for ODD and environment forecasting are initialized using prior values, and an update is performed after the second cell is observed. Finally, $p(\theta_1 \vert D_1)$ is computed.
    
    \item For cells $t=2$ until $t=N$, the following steps are repeated.
    
    \begin{enumerate}
        \item The ADS enters cell $t$ in AUTON mode. 
        \item The Dirichlet distributions are updated as cell $t+1$ is observed.
        \item $p(\theta_t \vert D_t)$ is updated. 
        \item Forecasts of obstacles and driver states for the next five cells are produced. 
        \item Trajectory planning is updated deciding the speed for the current cell and the five subsequent cells. Decisions are transmitted immediately. The utility perceived in the current cell is computed.
        \item If the probability of finding a puddle in cells $t+2$ or $t+3$ is greater than $q_{p}=0.25$, an alarm indicating a dangerous environmental state is raised. Similarly, if the probability of finding a puddle or a rock in cells $t+4$ or $t+5$ is greater than $q_p=0.25$ and $q_r=0.15$, respectively, another alarm is raised.
        \item If $p(\theta_{t+k} = \text{dist} \vert D_t)$ for $k=1,2,\dots, 5$ is above 0.9, a bad driver state alarm is issued.
        \item When a dangerous road state alarm is raised, driving modes are assessed as in Algorithm \ref{alg:auton}. The expected utility of each mode is computed within the next 5 cells. If MANUAL mode is preferred over AUTON, an RtI is issued.
        \item When issuing the RtI, if the probability of the driver being distracted is above 0.9, an emergency alarm is raised, and AUTON mode continues.
        If this probability is above 0.5 but below 0.9, a warning alarm is raised; however, the mode is changed to MANUAL. If this probability is below 0.5, the driving mode is changed to MANUAL without an alarm.
        \item The driving mode remains in MANUAL until the probability of the driver being distracted surpasses $q_d = 0.75$. If this does not happen within 10 cells, the AUTON mode is restored. 
        
    \end{enumerate}
    
    \item Upon arriving at cell $t=N$, the trip concludes and the globally attained utility is computed.

\end{enumerate}

\subsection{Results}

We implemented the previous scheme in Python. A class was defined with five main methods: \textit{predict} provides relevant forecasts at each cell for the subsequent 5 cells; the \textit{update} method updates current estimates of the driver and environment states given the newly observed information; \textit{decide} outputs the decision for the current cell and trajectory plan for the next 5 cells; \textit{issue warnings} 
evaluates the driver and environment state forecasts, emitting warnings based on them; and, finally, \textit{evaluate driving modes}
 evaluates the AUTON and MANUAL modes and decides whether to issue an RtI. 

An additional \textit{move} method coordinates  the previous ones:
when invoked, the ADS advances one cell,  \textit{update} is used to process new information, \textit{predict} produces relevant forecasts that are used by \textit{decide} and \textit{issue warnings}. Finally, if road state warnings are produced, \textit{evaluate driving modes} decides whether to issue a RtI and change the driving mode. 

\begin{figure}[h!] 
        \centering
        \includegraphics[width=0.8\textwidth]{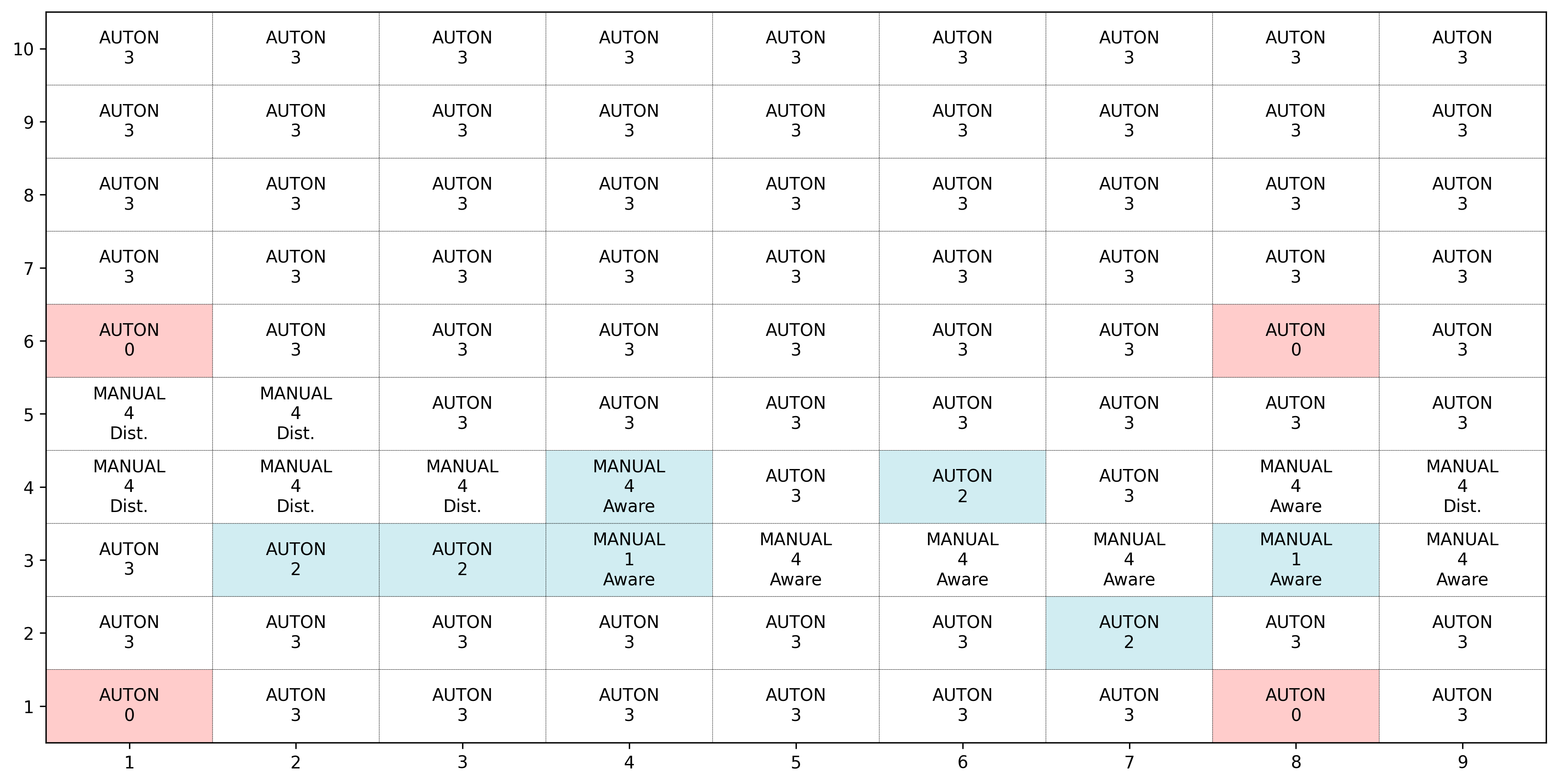}
        \caption[]
        {\small Mode and speed decisions in part of the road. } 
        \label{fig:road}
    \end{figure}

Figure \ref{fig:road} provides an example of a 90-cell road segment. To optimize space, we plot the one-dimensional road segment in two dimensions. The car beings its journey in the bottom left cell and moves within that row towards towards the right; once it reaches the right-most cell, it continues in the left-most cell of the row above. In each cell, we show the selected driving mode and speed. When in MANUAL mode, we show also whether the driver was aware or distracted. As can be seen in Figure \ref{fig:road}, the AUTON mode is clearly preferred over much of the journey. However, in row three column two, the presence of two consecutive puddles, as well as the fact that the driver is aware, triggers the MANUAL mode. After six additional cells, the driver starts to be distracted (i.e., row four column one). A puddle appears in row four column four, and the car skids as it  crosses at maximum speed. Note that the decision made in the cell containing the puddle was taken in the previous cell when the driver was distracted. Subsequently, due to the presence of another puddle, MANUAL mode is again invoked in row four column eight. However, as in the following cells, the driver is detected to be distracted and the AUTON mode is quickly restored.

The subsequent experiments highlight that the management scheme proposed in Section \ref{secMgmt} is effective by performing computational tests in our simulated environment. To that end, we perform three blocks of experiments. The first one  illustrates that the scheme produces reasonable outputs. The second evaluates the accuracy of inferences on the road and driver state produced by the ADS. The final block depicts how the outputs change under disparate environmental conditions. All code to reproduce the experiments is available at \url{https://github.com/roinaveiro/ads_trusto}.

\begin{figure}[h!] 
        \centering
        \begin{subfigure}[b]{0.475\textwidth}
            \centering
            \includegraphics[width=0.9\textwidth]{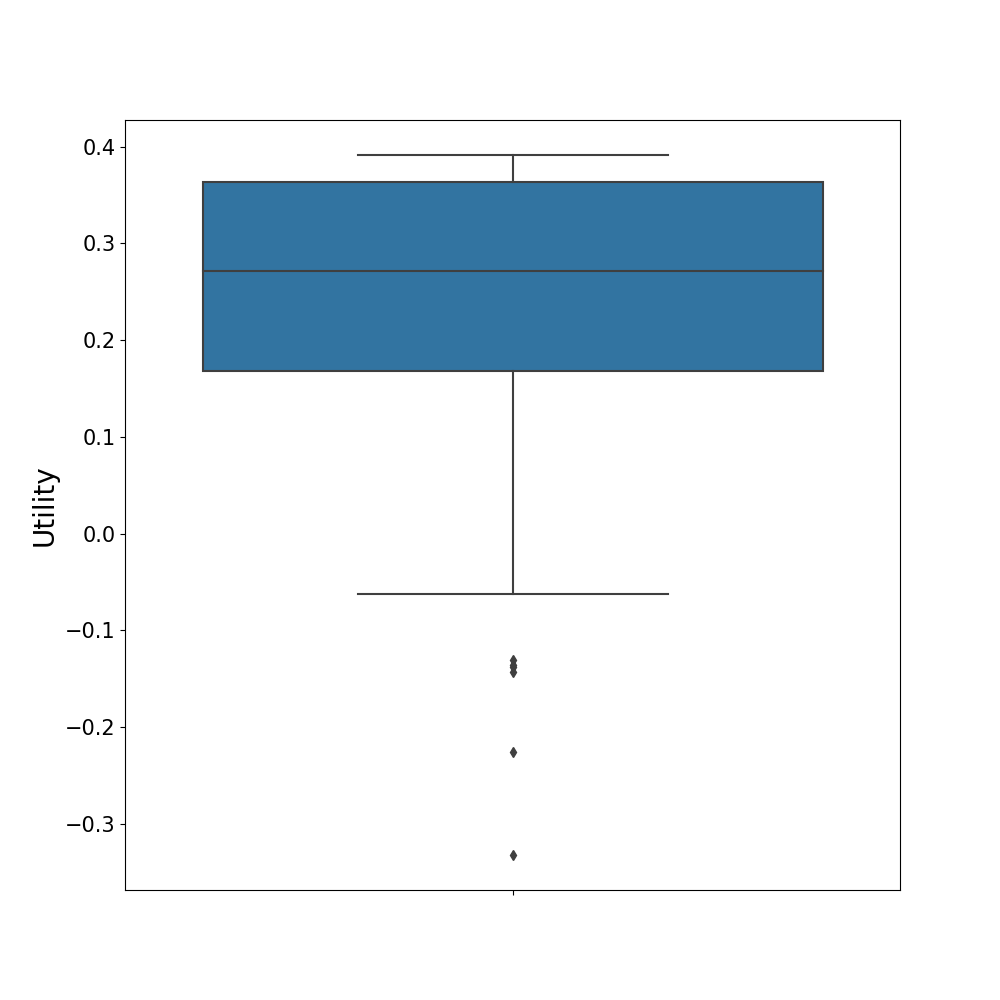}
            \caption[]%
            {{\small Utility perceived in the trip}}    
            \label{fig:mean and std of net14}
        \end{subfigure}
        \hfill
        \begin{subfigure}[b]{0.475\textwidth}  
            \centering 
            \includegraphics[width=0.9\textwidth]{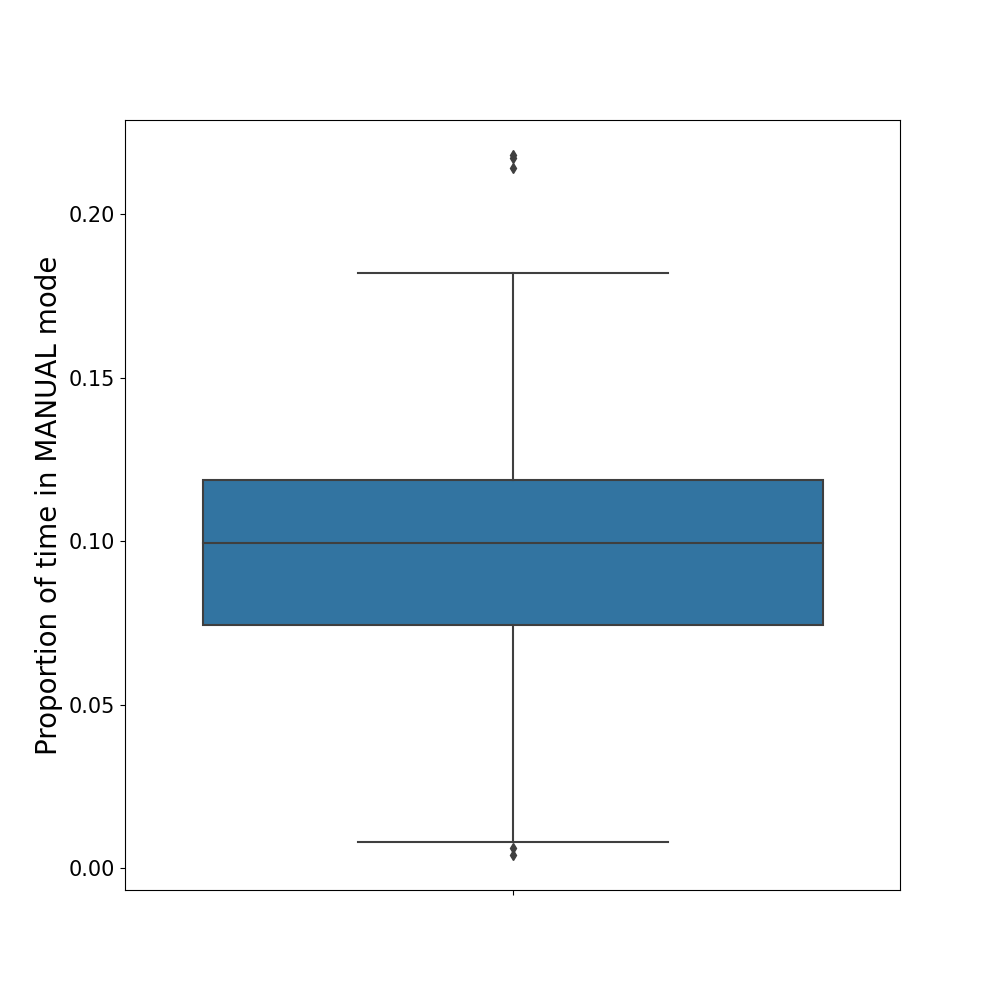}
            \caption[]%
            {{\small Proportion of time in MANUAL mode}}    
            \label{fig:mean and std of net24}
        \end{subfigure}
        \vskip\baselineskip
        \begin{subfigure}[b]{0.475\textwidth}   
            \centering 
            \includegraphics[width=0.9\textwidth]{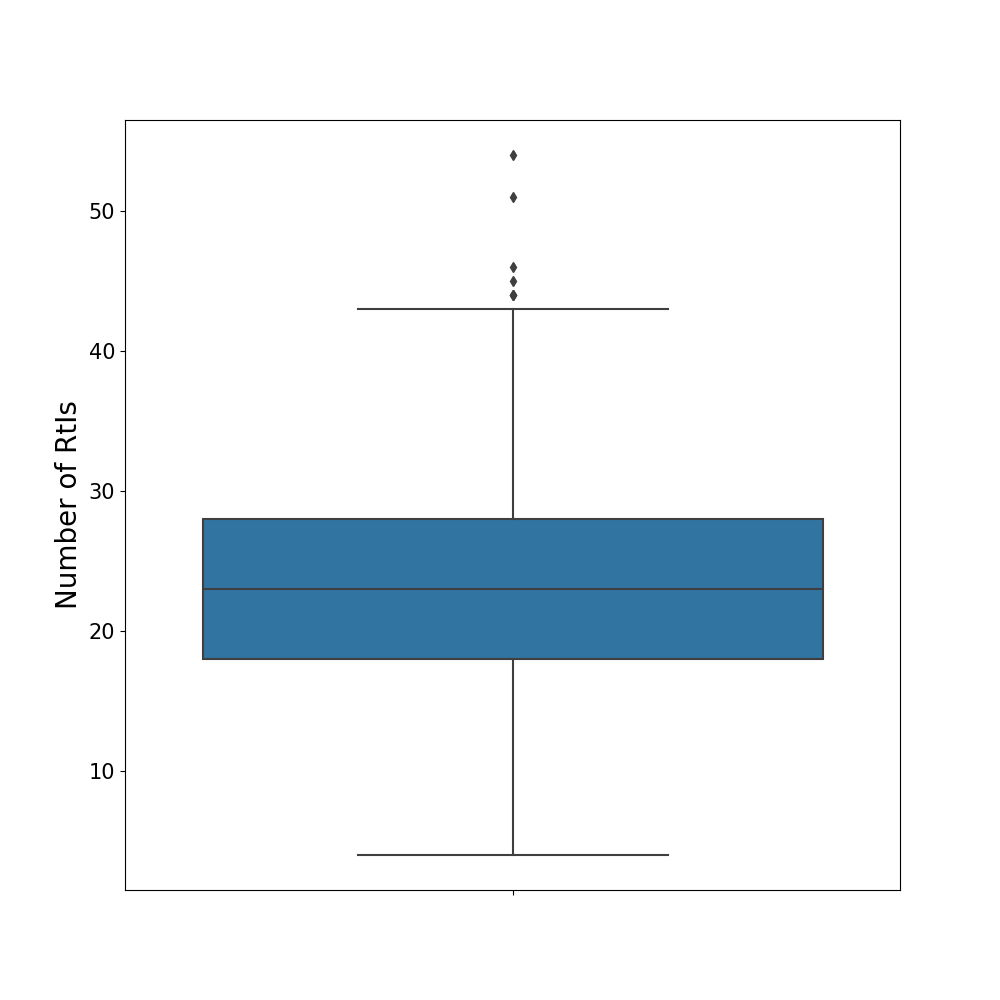}
            \caption[]%
            {{\small Number of RtIs issued}}    
            \label{fig:mean and std of net34}
        \end{subfigure}
        \hfill
        \begin{subfigure}[b]{0.475\textwidth}   
            \centering 
            \includegraphics[width=0.9\textwidth]{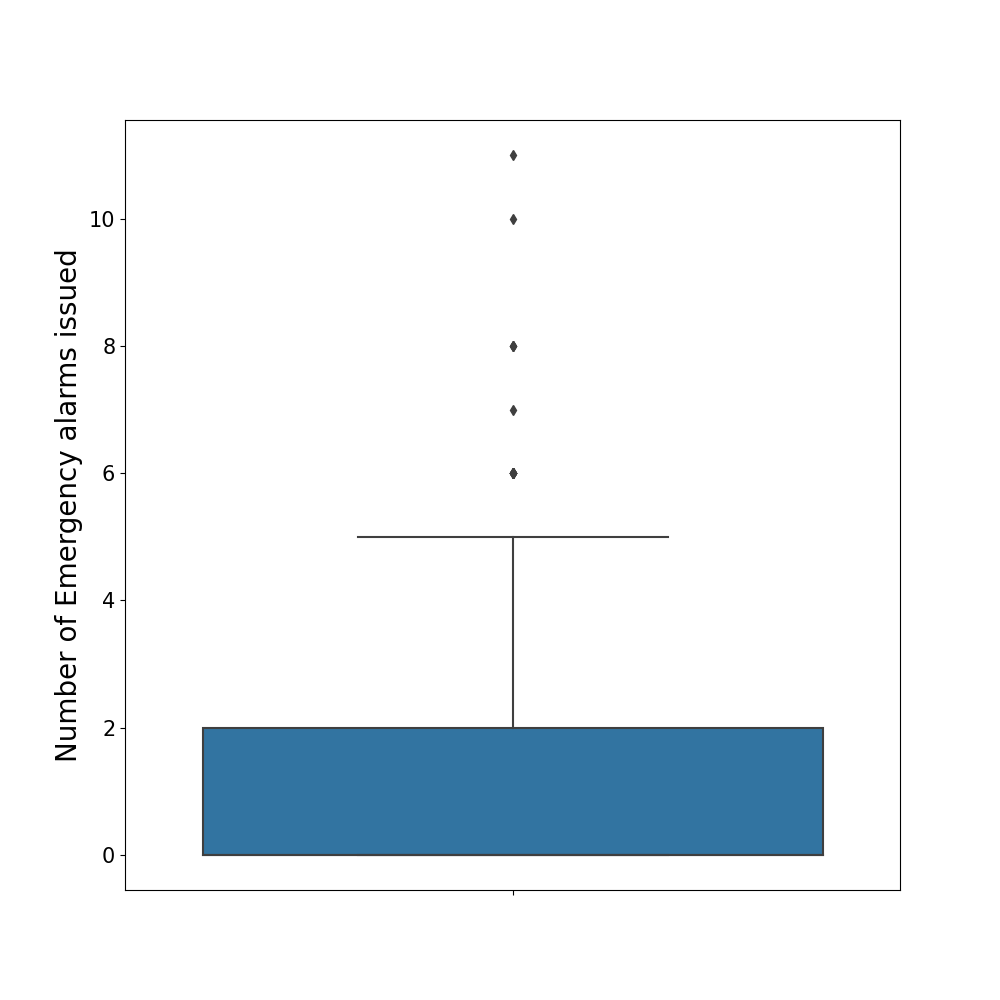}
            \caption[]%
            {{\small Number of emergency alarms issued}}    
            \label{fig:mean and std of net44}
        \end{subfigure}
        \caption[]
        {\small Box plots of relevant outputs I. } 
        \label{fig:block1_1}
    \end{figure}

For the first block, we perform 1000 simulations producing a 1000-cell road on each experimental run. The box plots in Figure \ref{fig:block1_1} depict the utilities attained, fraction of time in autonomous mode, number of RtIs issued and number of emergencies encountered. It can be observed in  Figure \ref{fig:block1_1}b
that, for the majority of simulated time, the ADS is in AUTON mode. Moreover, based on the frequencies of RtIs and emergencies (Figs
\ref{fig:block1_1}c and d), 
it can be inferred that the simulated environment 
is relatively dangerous. It is important to test ADS algorithms under such conditions as they represent worst-case scenarios with regard to safety.  Observe, as a consequence, 
the low utilities attained (i.e., Figure \ref{fig:block1_1}a). 
    
\begin{figure}[h!] 
        \centering
        \begin{subfigure}[b]{0.475\textwidth}
            \centering
            \includegraphics[width=\textwidth]{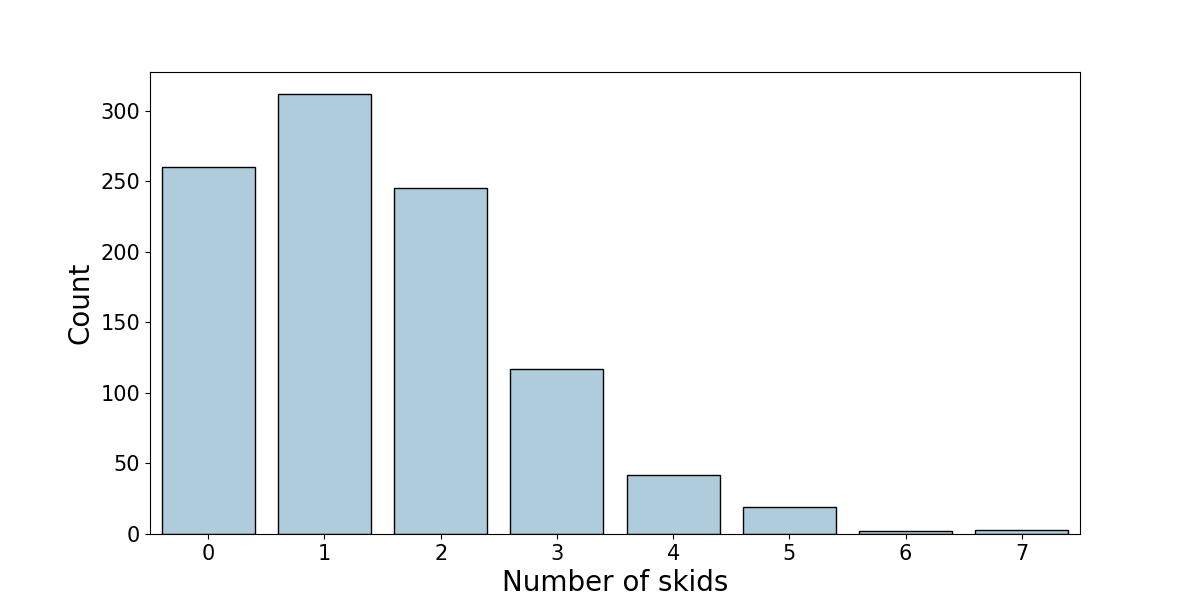}
            \caption[]%
            {{\small Number of skids}}    
            \label{fig:mean and std of net14}
        \end{subfigure}
        \hfill
        \begin{subfigure}[b]{0.475\textwidth}  
            \centering 
            \includegraphics[width=\textwidth]{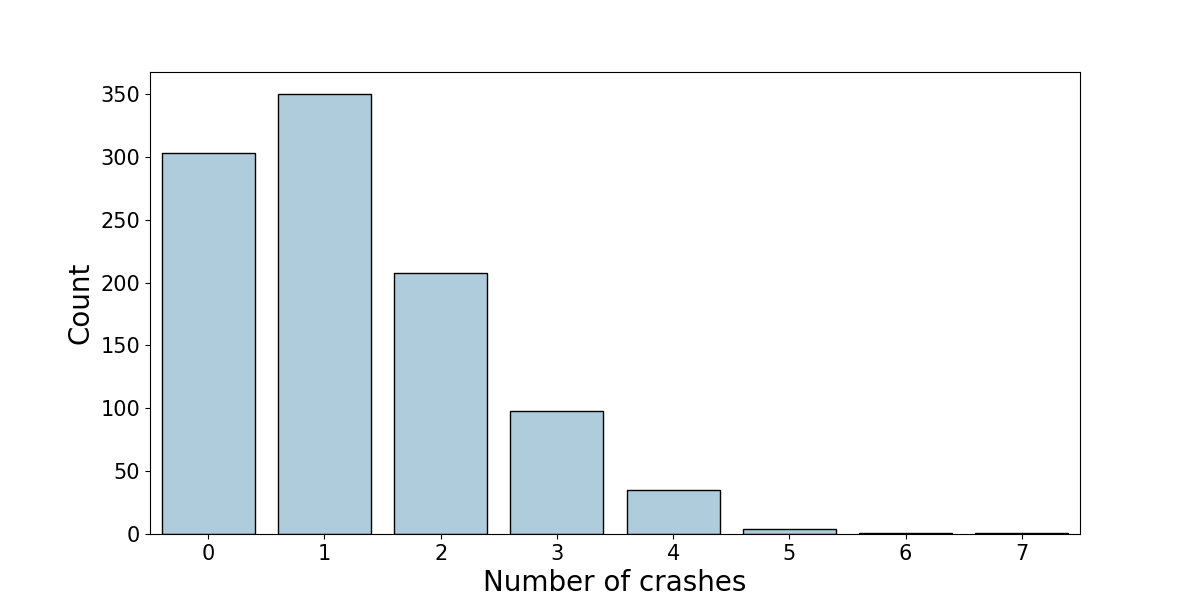}
            \caption[]%
            {{\small Number of crashes}}    
            \label{fig:mean and std of net24}
        \end{subfigure}
        \vskip\baselineskip
        \begin{subfigure}[b]{0.475\textwidth}   
            \centering 
            \includegraphics[width=\textwidth]{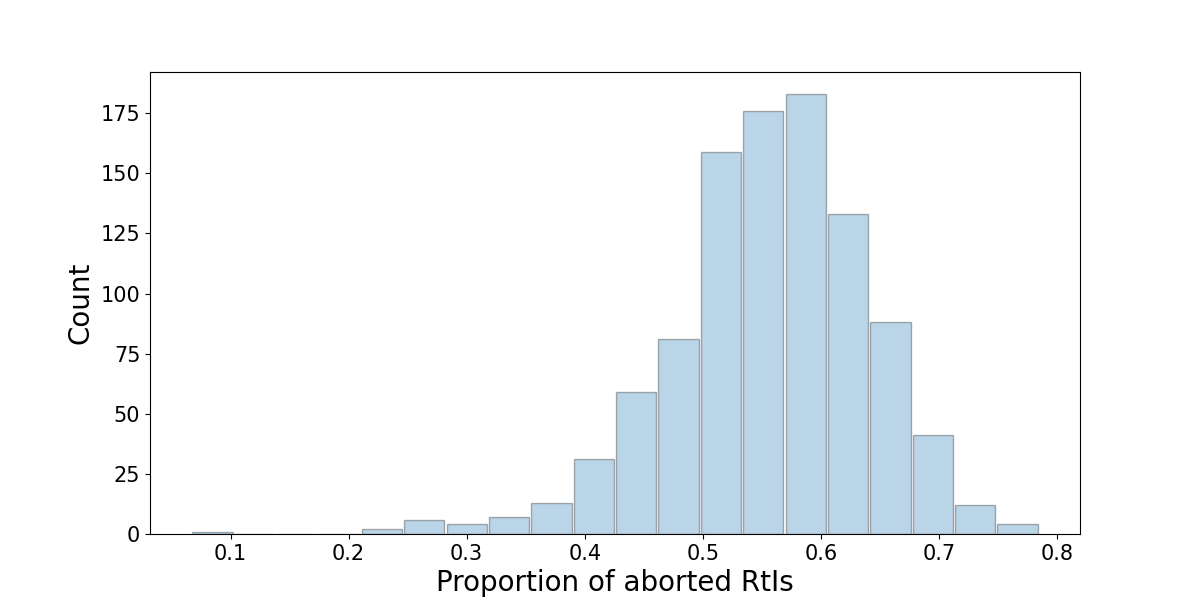}   
            \caption[]%
            {{\small Proportion of aborted RtIs}}    
            \label{fig:mean and std of net34}
        \end{subfigure}
        \hfill
        \begin{subfigure}[b]{0.475\textwidth}   
            \centering 
            \includegraphics[width=\textwidth]{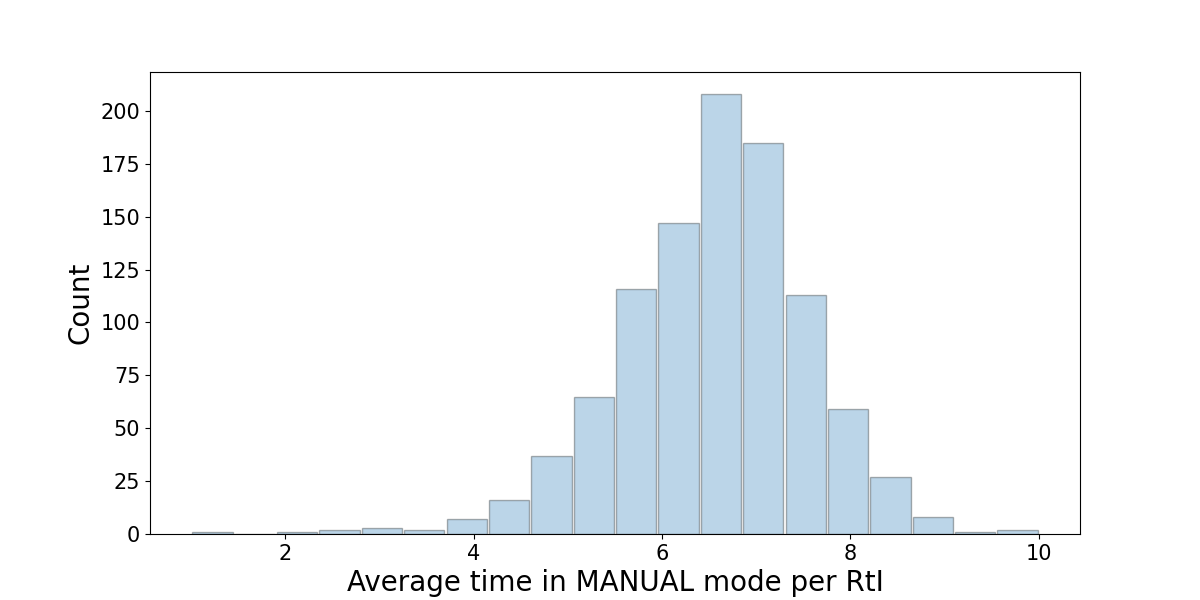}
            \caption[]%
            {{\small Average length in MANUAL mode per RtI}}    
            \label{fig:mean and std of net44}
        \end{subfigure}
        \caption[]
        {\small Histogram of relevant outputs I.} 
        \label{fig:block1_2}
    \end{figure}

Additionally, histograms on the number of skids, crashes, proportion of aborted RtIs, and average length in MANUAL mode per RtI are depicted in Figure \ref{fig:block1_2}. An aborted RtI is characterized by a high probability of surpassing ODD limits but wherein AUTON yields greater utility than MANUAL; since no EMERGENCY mode has been included, the ADS just stops. This output illustrates that the simulated environment in which the ADS operates is perilous. More often than not, the ADS encounters some adverse safety situation (i.e., a skid or a crash). However, it must be noted that, for all instances involving a crash, the vehicle is being operated in MANUAL mode with a distracted driver when the crash occurs. Therefore, the observed accidents are not induced by the self-driving features but by human error. 

\begin{figure}[h!] 
        \centering
        \begin{subfigure}[b]{0.475\textwidth}
            \centering
            \includegraphics[width=0.9\textwidth]{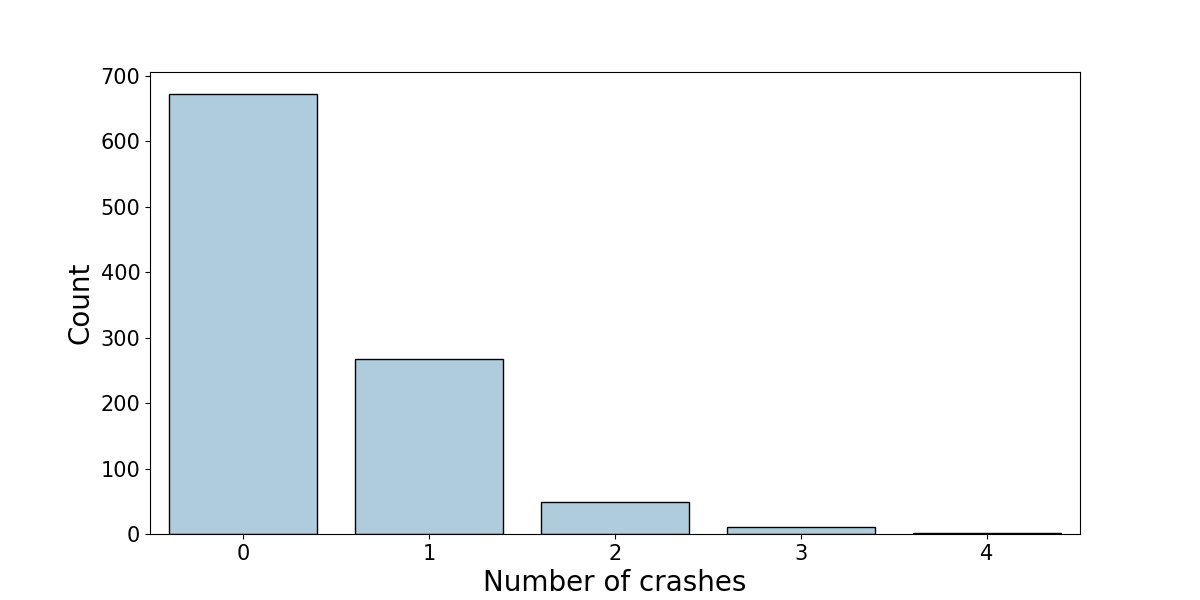}
            \caption[]%
            {{\small Number of crashes.}}
        \end{subfigure}
        \hfill
        \begin{subfigure}[b]{0.475\textwidth}  
            \centering 
            \includegraphics[width=0.9\textwidth]{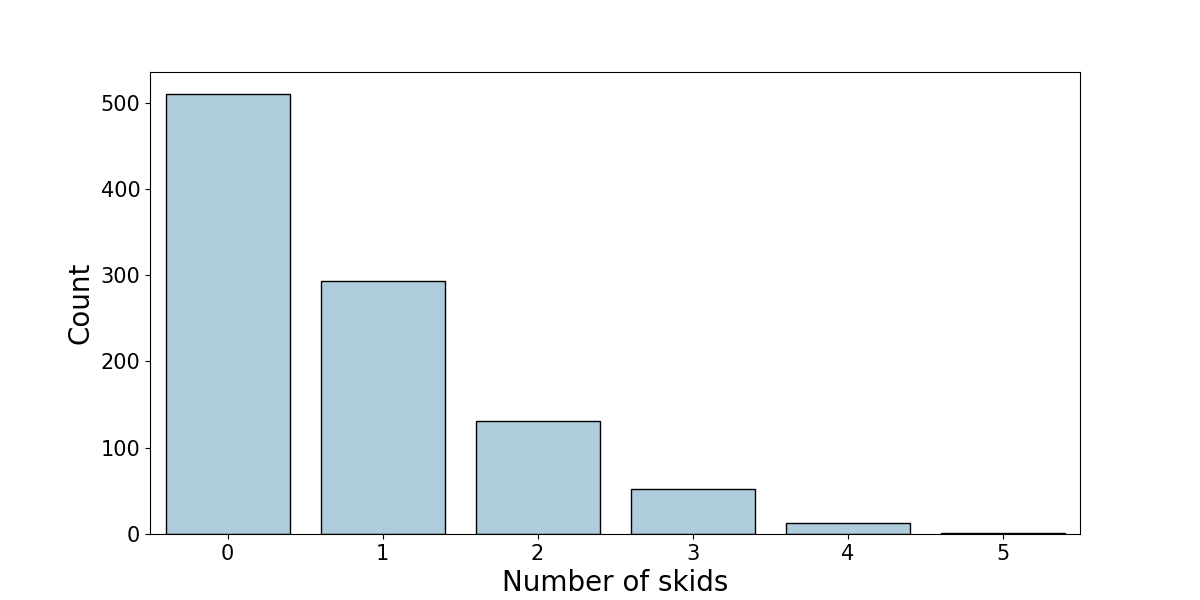}
            \caption[]%
            {{\small Number of skids.}}    
        \end{subfigure}
        \caption[]%
        {\small Histograms of relevant outputs II. } 
        \label{fig:less}
    \end{figure}

\begin{figure}[h!] 
        \centering
        \begin{subfigure}[b]{0.475\textwidth}
            \centering
            \includegraphics[width=0.9\textwidth]{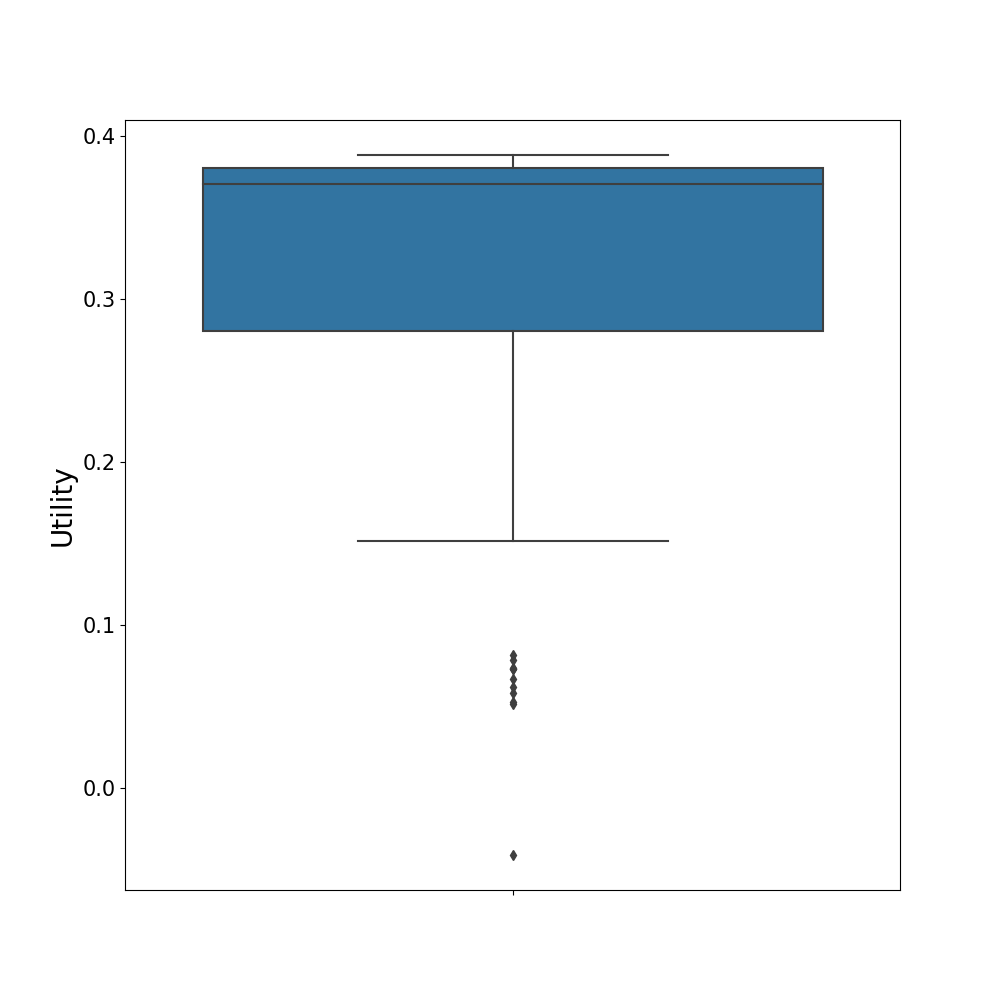}
            \caption[]%
            {{\small Utility perceived in the trip.}}
            \label{ut_less}
        \end{subfigure}
        \hfill
        \begin{subfigure}[b]{0.475\textwidth}  
            \centering 
            \includegraphics[width=0.9\textwidth]{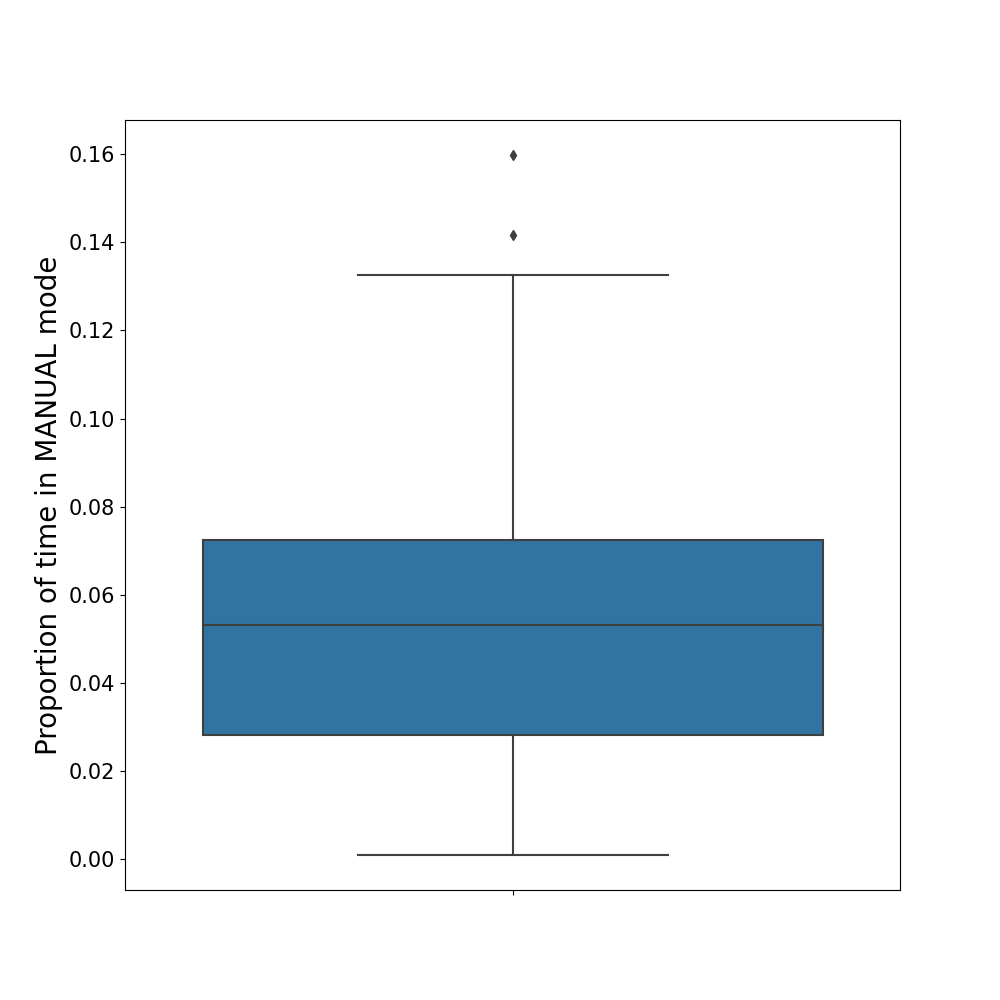}
            \caption[]%
            {{\small Proportion of time in MANUAL mode.}}   
            \label{prop_man_less}
        \end{subfigure}
        \caption[]%
        {\small Boxplot of relevant outputs II. } 
        \label{fig:less_bp}
    \end{figure}

Indeed, we can check this, by altering the conditions necessitating the issue of an RtI and by requiring that the driver maintain control once a MANUAL mode transition occurs.
Figure \ref{fig:less} shows the number of crashes and skids in another experiment with greater values of $q_r$ and $q_p$, and a lesser value of $q_d$ (0.25, 0.3, and 0.5, respectively). It can be observed that substantially less skids and crashes occur. This is a consequence of decreasing the time in MANUAL mode, as illustrated in Figure \ref{prop_man_less}; the corresponding gain in utility can be seen in Figure \ref{ut_less}.

\begin{figure}[h!] 
        \centering
        \begin{subfigure}[b]{0.475\textwidth}
            \centering
            \includegraphics[width=0.9\textwidth]{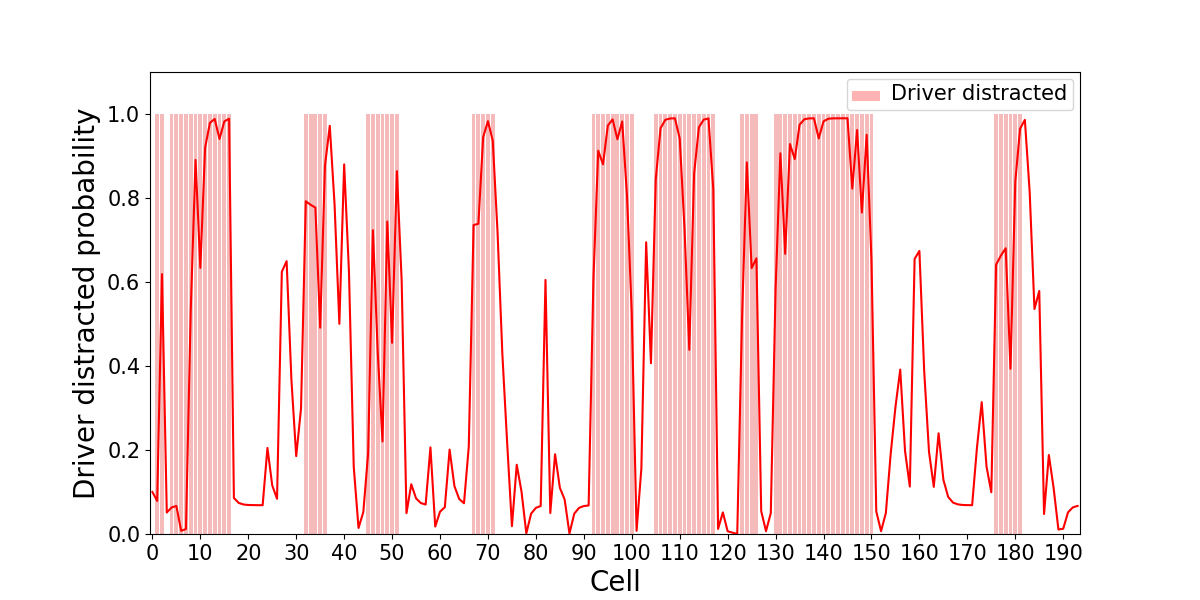}
            \caption[]%
            {{\small Estimate of the probability of driver distracted and true driver state.}}
            \label{ddist}
        \end{subfigure}
        \hfill
        \begin{subfigure}[b]{0.475\textwidth}  
            \centering 
            \includegraphics[width=0.9\textwidth]{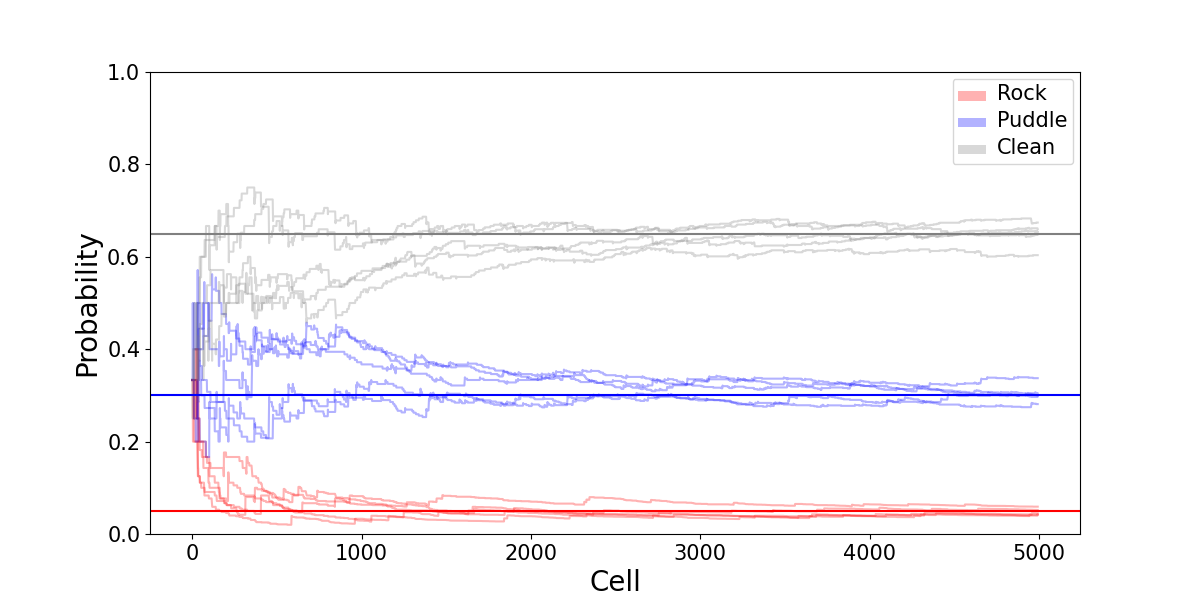}
            \caption[]%
            {{\small Estimate of the probability of each obstacle in cell $t$ given that cell $t-1$ contained a puddle.}}    
            \label{env}
        \end{subfigure}
        \caption[]%
        {\small Accuracy of inferences. } 
        \label{fig:acc_inf}
    \end{figure}

Figure \ref{fig:acc_inf} seems to validate the accuracy of the inferences produced by the ADS. Figure \ref{ddist} represents the estimate $p(\theta_t = \text{dist} \vert D_t)$ through time. The presence of a vertical red bar at a given time, indicates that the driver state is distracted. Correspondingly, 
observe that the estimated  probability of being distracted
tends to be higher when the driver is actually distracted. Moreover, Figure \ref{env} represents the estimated probability of cell $t$ containing a rock, a puddle or being clean, given that cell $t-1$ contained a puddle. We plot the estimates associated with five different simulations. The horizontal lines represent the true values of these probabilities (Table \ref{kakaroad}). It can be observed that the estimates of these quantities converge to the true probability as the ADS proceeds through time.

A natural question to ask relates to how the simulated ADS results change based on varying the environment conditions presented in Table \ref{kakaroad}. The third block of experiments provides some clarity to this question. More specifically, we change the proportion of puddles and observe how the simulated ADS behavior changes. The proportion of puddles is altered by varying the value of $p(y_{t+1} = \text{puddle} \vert y_{t} = \text{clean} )$ from Table \ref{kakaroad}. The influence of the proportion of rocks on the simulation outputs has been also measured but is not presented due to lack of space. Results were qualitatively similar. 

\begin{figure}[h!] 
        \centering
        \begin{subfigure}[b]{0.475\textwidth}
            \centering
            \includegraphics[width=0.9\textwidth]{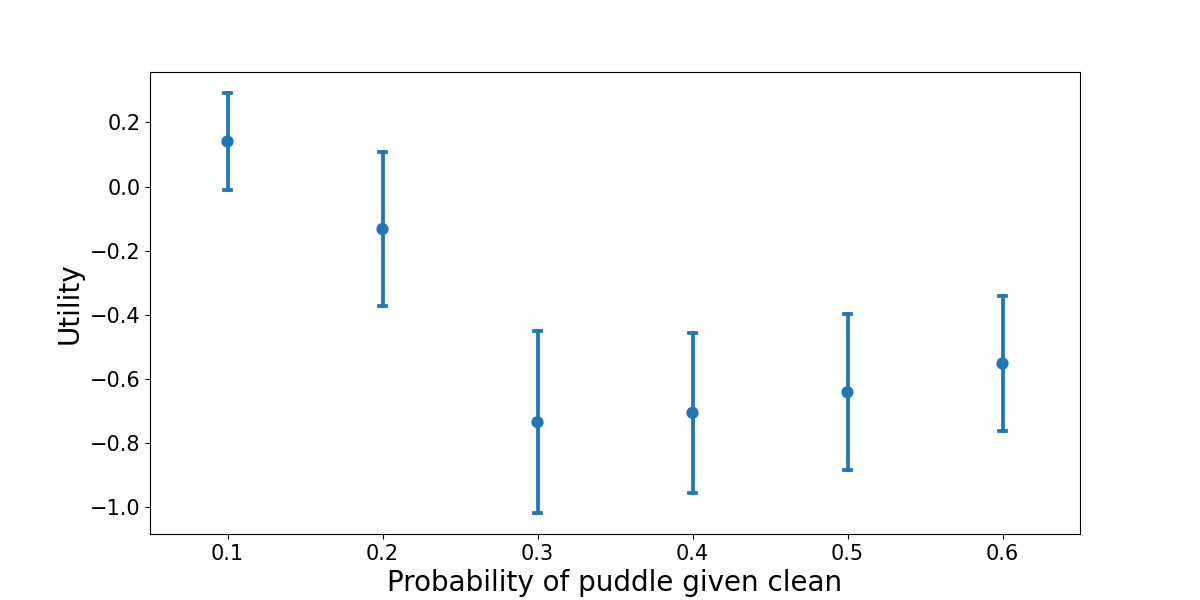}
            \caption[]%
            {{\small Utility perceived for different values of $p(y_{t+1} = \text{puddle} \vert y_{t} = \text{clean} )$.}}
            \label{ut_pudd}
        \end{subfigure}
        \hfill
        \begin{subfigure}[b]{0.475\textwidth}  
            \centering 
            \includegraphics[width=0.9\textwidth]{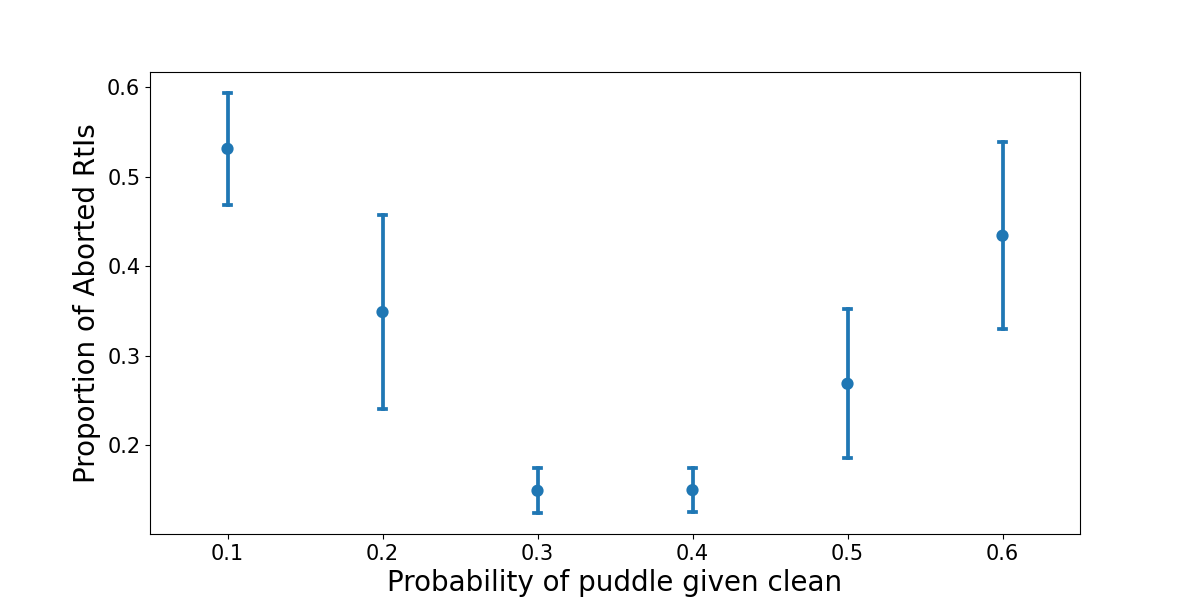}
            \caption[]%
             {{\small Proportion of aborted RtIs for different values of $p(y_{t+1} = \text{puddle} \vert y_{t} = \text{clean} )$.}}
            \label{rej_pudd}
        \end{subfigure}
        \caption[]%
        {\small Study for different values of different values of $\bm{ p(y_{t+1} = \text{puddle} \vert y_{t} = \text{clean} ) }$. } 
        \label{fig:pudd}
    \end{figure}

In particular, we simulated for six different values: 0.1, 0.2, \dots, 0.6. For each puddle proportion, we performed 100 simulations with a 1000 cells road. Average utility attained is presented in Figure \ref{ut_pudd}. Error bars correspond to a single standard deviation. Herein, we observe an interesting phenomenon. As may be expected, increasing the proportion of puddles tends to produce a decrease in average utility. When more puddles are expected, the number of RtIs increases. Furthermore, when there are many puddles, the probability of the driver being aware is much higher, i.e., see Table \ref{kakaware}. Therefore, for large values of $p(y_{t+1} = \text{puddle} \vert y_{t} = \text{clean} )$ it is less likely to encounter a distracted driver. Thus, when evaluating driving modes as a consequence of a potential risk, it is more likely that MANUAL is better than AUTON, making the transition to MANUAL more likely. However, as the driver is asked to intervene more often, the collective risk of a distracted driver controlling the vehicle increases, thereby decreasing average utility.

Conversely, when increasing the proportion of puddles beyond a certain limit, we observe that the utility starts to increase. The reason for this phenomenon is that, when many puddles are expected, the ADS detects driver underperformance
more often. In turn, the ADS does not allow them to take control over the vehicle as frequently. This behavior can be observed in Figure \ref{rej_pudd} wherein we see a decrease in the risk of a distracted driver in control, thereby increasing utility. 

The observed behavior in Figure \ref{fig:pudd} is an incarnation of what we previously referred  to as a \textit{fundamental dilemma} in level-3 and -4 ADS. Such questions do not have simple
answers, but we have illustrated a potential resolution to the dilemma herein by maximizing the expected utilities defined in Table \ref{kakaspeed}.


\section{Discussion} \label{secDiscussion}
We have provided a global Bayesian decision-making model to support the management of driving modes in level-3 and -4 ADS. The integrated approach takes into account the operational design domain, dynamic driver state modeling, environment monitoring, and  driver intervention performance assessment within an expected utility framework. It serves for managing driving modes and issuing early warnings.

Our proposed DIPA framework can be extended to consider driver reaction time. By combining information from the driver state forecast with the statistical learning process for driver intervention performance, the speed with which a driver reacts to an RtI can be anticipated and, potentially improved, after repeatedly sampling RtI decisions. 
Additionally, machine learning algorithms can be utilized to monitor driver performance to acquire an estimate of their driving aptitude. That is, RtIs can be executed in non-hazardous situations to perform DIPAs to determine such aptitude. Moreover, individual differences in aptitude may affect the degree to which environmental factors affect their alertness, implying that alternative Bayesian forecasting models may prove more accurate than others. Future research may thus include competing forecasts models evaluated for predictive accuracy using the model monitoring methods of \citet{westharrison}. Competing forecast models for the environmental variables may similarly improve the performance of the methods presented herein, especially for ADS that operate in multiple, distinct environments (e.g., semi-trailer trucks crossing international borders). 

With regards to ethical decision making, a universal consensus of an ADS's programmed morality is unlikely. 
Since our driving mode management framework is rooted in decision theory, the selected utility function is a critical factor of the ADS' performance. Despite criticism of trolley problems in the literature, their study provides the ADS with a consequentialist ethical perspective by informing the objective function's structure; however, such research must be approached with care. The crowd-sourced nature of {\em moral-machine-like} experiments limits the effect of individual subjectivity, but it also risks overwhelming opinions of minority groups. Ultimately, an ADS utility function can adopt either an egalitarian (maximizing social welfare) or a self-preserving (maximizing driver's survival) perspective. The degree to which a populace prefers one perspective to another is likely dependent on a myriad of sociological and cultural factors. Therefore, it will be useful to develop a suite of generic parametric utility models for ADS management from which designers, owners and policymakers could choose, similar to the  proposal by \cite{defendercase} in cybersecurity. Thus, one could incorporate ethics into the utility functions to make moral tradeoffs computationally tractable 

Finally, aside from the utility function, the algorithms developed herein rely upon probabilistic thresholds to transition between driving modes and issue warnings. These thresholds are highly influential with regards to ADS performance and help determine the amount of collective risk accepted. In this paper, these values are treated as exogenously determined parameters; however, future research may seek to optimize them. Such a perspective provides a next step in the current line of research via the collective optimization of thresholds for interacting autonomous and manned vehicles. The application of both reinforcement learning and approximate dynamic programming hold great promise in this regard.

\section*{Disclaimer}The views expressed in this article are those of the authors and do not reflect the official policy or position of the United States Air Force, United States Department of Defense, or United States Government.

\section*{Acknowledgments}
This work has been performed in the context of the Trustonomy project, which has received funding from the European Community’s Horizon 2020 research and innovation programme under grant agreement No 812003, an 
NSF grant DMS-1638521 at SAMSI and a BBVA Foundation project.
DRI is supported by the AXA-ICMAT Chair and the Spanish Ministry of Science program MTM2017-86875-C3-1-R.  
RN also acknowledges support of the Spanish Ministry for his grant FPU15-03636.

%

%
%

\clearpage

\bibliographystyle{informs2014trsc} 
\bibliography{RtI_biblio.bib} 


\end{document}